\newif\ifspin
\spinfalse

\documentclass[aps,floatfix,twocolumn,showpacs,10pt,nofootinbib]{revtex4-1}
\usepackage[T1]{fontenc}
\usepackage{amsfonts}
\usepackage{hyperref}
\usepackage[usenames,dvipsnames]{color}
\newcommand{\be}{\begin{equation}}
\newcommand{\ee}{\end{equation}}
\newcommand{\mysubsection}[1]{ \vspace{0.3cm} \begin{center}\textbf{#1} \vspace{0.3cm}\end{center}}
\setcitestyle{super}			

\usepackage{amssymb,amsmath,graphicx}

\begin{document}


\title{g-factor of electrons in gate-defined quantum dots in a strong in-plane magnetic field}


\author{Peter Stano$^{1,2,3}$, Chen-Hsuan Hsu$^1$, Marcel Serina$^4$, Leon C. Camenzind$^4$, Dominik M. Zumb\"uhl$^4$, and Daniel Loss$^{1,4}$}
\affiliation{$^{1}$RIKEN Center for Emergent Matter Science (CEMS), Wako, Saitama 351-0198, Japan}
\affiliation{$^{2}$Department of Applied Physics, School of Engineering, University of Tokyo, 7-3-1 Hongo, Bunkyo-ku, Tokyo 113-8656, Japan}
\affiliation{$^{3}$Institute of Physics, Slovak Academy of Sciences, 845 11 Bratislava, Slovakia}
\affiliation{$^{4}$Department of Physics, University of Basel, Klingelbergstrasse 82, CH-4056 Basel, Switzerland}


\date{\today}

\begin{abstract}
We analyze orbital effects of an in-plane magnetic field on the spin structure of states of a gated quantum dot based in a two-dimensional electron gas. Starting with a $k \cdot p$ Hamiltonian, we perturbatively calculate these effects for the conduction band of GaAs, up to the third power of the magnetic field. We quantify several corrections to the g-tensor and reveal their relative importance. We find that for typical parameters, the Rashba spin-orbit term and the isotropic term, $H_{43} \propto {\bf P}^2 {\bf B} \cdot \boldsymbol{\sigma}$, give the largest contributions in magnitude. The in-plane anisotropy of the g-factor is, on the other hand, dominated by the Dresselhaus spin-orbit term. At zero magnetic field, the total correction to the g-factor is typically 5-10\% of its bulk value. In strong in-plane magnetic fields, the corrections are modified appreciably.
\end{abstract}


\maketitle

\newcommand{\x}{\hat{\textbf{x}}}
\newcommand{\y}{\hat{\textbf{y}}}
\newcommand{\z}{\hat{\textbf{z}}}

\newcommand{\Bin}{\textbf{b}}
\newcommand{\Binscalar}{b}
\newcommand{\Bz}{B_z}
\newcommand{\Bzvec}{B_z \z}
\newcommand{\Rin}{\textbf{r}}
\newcommand{\Rz}{z\z}
\newcommand{\Pin}{\textbf{p}}
\newcommand{\Ain}{\textbf{a}_{||}}
\newcommand{\Ainx}{a_{||,x}}
\newcommand{\Ainy}{a_{||,y}}
\newcommand{\Az}{\textbf{a}_{\pepr}}

\newcommand{\ave}[2]{\overline{#1}^{#2}}

\newcommand{\flux}{\Phi}

 \section{Introduction}

Spin qubits in gated quantum dots\cite{loss1999,hanson2007,kloeffel2013} based in two dimensional electron gas (2DEG) are now seeing a resurge in interest due to a recent progress in GaAs\cite{delbecq2016,nakajima2017,hoffman2017,botzem2016,nichol2017,malinowski2017} and, especially, in potentially nuclear-spin-free materials like Si,\cite{yoneda2017,watson2017,maurand2016,reed2016, laucht2017,harvey-collard2017,zajac2016,yang2016} Ge,\cite{hendrickx2018:A,watzinger2018:A} and C.\cite{eich2018} Many of the experiments are done applying relatively strong in-plane magnetic fields, in the order of Teslas. It is a well established fact that such in-plane magnetic fields have sizable effects in 2DEGs.\cite{stern1968} This motivated us in Ref.~\citenum{stano2017} to analyze the effects of the in-plane magnetic fields on the orbital structure of the quantum dot states. There, we laid down the theory for using such effects as a new spectroscopic tool of quantum dots. The idea was conceived in Ref.~\citenum{camenzind2}, which demonstrated that the shape of quantum mechanical orbitals of a quantum dot can be inferred in this way. The information on the quantum dot shape thus acquired was essential for the experimental quantification of the spin-orbit couplings in Ref.~\citenum{camenzind1}, further demonstrating the power of this tool.

In this article, we extend the investigations of Ref.~\citenum{stano2017} to the spin structure of a quantum dot. 
The spin-dependent corrections due to the orbital effects of the in-plane field appear, first, as corrections to the spin-orbit interactions, such as Rashba and Dresselhaus terms in GaAs.
Importantly, in the presence of magnetic field additional spin-orbit terms arise, which are present even in bulk- and interface- inversion symmetric structures. These, as well as the magnetic-field induced corrections to the inversion-asymmetry originated ones, are not time reversal symmetric.  
They can therefore directly---in the lowest order---change the energy splitting of a pair of time reversed states (spin `up' and `down' corresponding to the same orbital). We expect that such energy effects are their most important consequence, and therefore mainly restrict ourselves to evaluating the corresponding renormalization of the g-factor. 

We derive a dozen of different terms for the g-factor corrections, Eqs.~\eqref{eq:gd}--\eqref{eq:gz00}, constituting our main results. They  differ in the dependence on the 2DEG width (increase or decrease), magnetic field magnitude (constant, or magnetic-field dependent), and direction (isotropic, anisotropic and relating to the crystal axes, or anisotropic and relating to the quantum dot axes), heterostructure interface electric field (dependent, or largely independent on it), and symmetry of the heterostructure confinement (present only in asymmetric 2DEGs or present also in symmetric quantum wells).

There is vast literature concerning g-factor theory and experiments. Instead of trying to give an overview, we only refer to works which have direct connection to our results. The g-factor corrections that we calculate here are solely bandstructure (or single particle) effects. They correspond to experiments with GaAs occupied by a single or a few particles.\cite{hanson2003,zumbuhl2004,kogan2004} In other words, our theory does not cover the g-factor changes arising from the electron--electron interaction-induced exchange,\cite{snelling1991,thomas1996} which is also modulated by magnetic field, for example, through the induced renormalization of the electron mass.\cite{tutuc2003} 
Second, we also do not analyze the effects of strain,\cite{rocca1988,kowalski1994} assuming that they are negligible in the lattice matched AlGaAs/GaAs heterostructres with the 2DEG relatively far below metallic surface gates. Finally, we focus on gated dots, where the effects are perturbative, unlike in self-assembled dots, where they are of order one.\cite{takahashi2010,takahashi2013} Among recent works, we point out Ref.~\citenum{michal2017} having partial overlap with what we do here,\footnote{The calculations done in Ref.~\citenum{michal2017} aim at explanations and fittings of the data of that particular measurement, rather than at a general g-factor theory.} and Ref.~\citenum{miserev2017b} focusing on holes and being similar in spirit.

Our results can be exploited in several ways. First, they should be taken as the theory accompanying the current experiments, which have in GaAs dots reached resolution required to extract effects of such small magnitude.\cite{camenzind3} Fitting data from such experiments, one could aim at extracting the $k \cdot p$ parameters,\footnote{For example, the g-factor variation with respect to the magnetic field in-plane direction reveals the bulk Dresselhaus constant, as discussed below and particularly in Fig.~\ref{fig:gs}f.}  which are still under vivid debate even in the best known semiconductors.\cite{devizorova2014} From the point of view of spin qubits, the inhomogeneities in the g-factor are a primary agent for, on one hand, spin manipulation and, on the other, coupling to the charge noise. 

The article is structured as follows. In Section II we present the approach. It is a perturbative calculation introduced in Ref.~\citenum{stano2017}. Here, we extend it by spin-dependent terms arising up to the fourth order in a $k \cdot p$ theory for the $\Gamma_6$ conduction band of a zinc-blende crystal with $T_d$ symmetry.
Section III exemplifies how the spin-dependent effects arise due to the in-plane field orbital effects, and motivates approximations which we adopt for the rest of the calculations. Section IV lists and analyzes the diagonal corrections to the g-tensor for a purely in-plane magnetic field. 
We defer calculational details and additional material to several appendices. Appendix \ref{app:off-diagonal} lists the off-diagonal components of the g-tensor. Appendix \ref{app:spin-dep} contains detailed derivations of all the g-tensor corrections using  third order perturbation theory. Appendix \ref{app:constant} lists dimensionless constants which enter the results. Appendix \ref{app:symmetric} discusses the g-factor corrections for a symmetric quantum well. Appendix \ref{app:B2} shows an example for the g-factor corrections quadratic in the magnetic field. Appendix \ref{app:lin-lin} estimates the leading correction of higher order in spin-orbit constants, showing that they are indeed negligible compared to the leading order ones that we discuss in the main text.

\section{Definitions and methods}

We now introduce the method. It is a straightforward extension of the approach explained in detail in Ref.~\citenum{stano2017}, so we only recapitulate it shortly. It starts with a three dimensional Hamiltonian of the heterostructure,
\be
H=
T(\textbf{P}) + V({\bf R}) + H_{\rm Z},
\label{eq:H zeroth order}
\ee
which comprises  the kinetic, potential, and Zeeman terms. The kinetic energy operator $T$ is a function of the kinetic momentum
\be
\textbf{P} = -i\hbar (\partial_x, \partial_y, \partial_z) + e\textbf{A},
\ee
where $e$ is the proton charge, and
the vector potential $\textbf{A}$ is due to the magnetic field $\textbf{B}=(B_x,B_y,B_z)$. The confinement potential $V(\textbf{R})$, is due to gates and material composition, as specified below. 

We consider a structure grown along a crystallographic axis, denoted by $\z \equiv [001]$, which we in further call the out-of-plane axis. The remaining two crystallographic axes are denoted by $\x \equiv [100]$
and $\y \equiv [010]$, and are called in-plane. With this notation, we set the unperturbed part of the three dimensional Hamiltonian as 
\be
H_0=h_z + h_{2D}.
\label{eq:H0}
\ee
It defines the basis for the perturbative calculations below. The unperturbed part is chosen separable in the in-plane and out-of-plane coordinates. Next, we describe these two parts in further detail.

\subsection{Unperturbed part defining the basis}

The unperturbed Hamiltonian for the heterostructure growth direction, along the unit vector $\z$, is\footnote{We stick here to the triangular heterostructure confinement in Eq.~\eqref{eq:Hz} and do not discuss in the main text, for the sake of brevity, other confinement types considered in Ref.~\citenum{stano2017}. We give results for a symmetric confinement in App.~\ref{app:symmetric}.}
\be
h_z=- \partial_z \frac{\hbar^2}{2m(z)} \partial_z + \Theta(z)e E_{\rm ext} z +\Theta(-z)V_0,
\label{eq:Hz}
\ee
where $\Theta(z)$ is the Heaviside step function, $V_0$ is the offset of the conduction bands of the constituent materials (we specify to Al$_{x}$Ga$_{1-x}$As, referred to as material $A$, and GaAs, referred to as material $B$), $E_{\rm ext}$ is the interface electric field, and the position-dependent effective mass is
\be
m(z) = \Theta(-z) m_A + \Theta(z) m_B.
\label{eq:mz}
\ee
 The spectrum of $h_z$ defines the subbands, denoted by $|\alpha\rangle$ with the corresponding energies $E_\alpha$. We use Greek indexes for subbands, with the ground state belonging to subband $\alpha=1$, while $\alpha =2$ is the lowest excited subband, and so on. 

Since the position dependence of the mass does not lead to spin-dependent effects, we approximate it by a constant within each subband, being $\ave{m(z)}{\alpha}$. The overline is defined as the average within the subband, 
\be
\ave{O}{\alpha} \equiv \langle \alpha | O | \alpha \rangle, \qquad \ave{O}{\alpha\beta} \equiv \langle \alpha | O | \beta \rangle,
\ee
and we also introduced the latter notation for further convenience. For the lowest subband, we set $\ave{m(z)}{\alpha} \approx m_B \equiv m$. In other words, even though we take the effects of mass inhomogeneity into account when constructing the basis, we do not include it among the considered perturbations.\footnote{The corrections resulting from such terms are expected to be much smaller than the terms denoted $g_z$ (see below), which are of similar origin and which are subdominant.} With that, we define the nominal width $l_z$ of the 2DEG by
\be
eE_{\rm ext} \equiv \hbar^2/2ml_z^3,
\label{eq:lz}
\ee
that is, $l_z$ is a quantity with the dimension of the length defined by the surface electric field and the effective mass.

The in-plane part of $H_0$, which defines the quantum dot, is taken with an anisotropic harmonic confinement
\be
h_{2D} = \frac{\Pin^2}{2m} + \frac{\hbar^2}{2m} \left( \frac{ x_d^2}{l_x^4} + \frac{y_d^2}{l_y^4} \right).
\label{eq:H2D}
\ee
Here, the confinement potential is expressed in the dot coordinates defined by unit vectors $\x_d$ and $\y_d$, which are rotated with respect to the crystallographic axes $\x$ and $\y$ by an angle $\delta$. The in-plane momentum contains the orbital effects due to the out-of-plane component of the magnetic field $\Bz$,
\be
\Pin= -i\hbar(\partial_x,\partial_y) + \frac{e \Bz}{2}(-y, x).
\label{eq:Pin}
\ee
The spectrum of $h_{2D}$ is equivalent to two independent linear harmonic oscillators with excitation energies $E_x$ and $E_y$. For $\Bz=0$ the two energies are given by $E_{x/y}=\hbar^2/ml_{x/y}^2$, while the symmetric case $l_x=l_y$ corresponds to the well known Fock-Darwin spectrum. The general case of $\Bz\neq 0$ and $l_x\neq l_y$ can  also be solved straightforwardly.\cite{rebane1969,schuh1985,davies1985} We use Roman indexes for the eigenstates of $h_{2D}$, called in-plane (orbital) states. We denote their wave functions by $|i\rangle$ and the corresponding energies by $E_i$. The two harmonic oscillators quantum numbers  corresponding to this state are denoted as $n_x^{(i)}$ and $n_y^{(i)}$. 

\begin{figure}
\includegraphics[width=1\columnwidth]{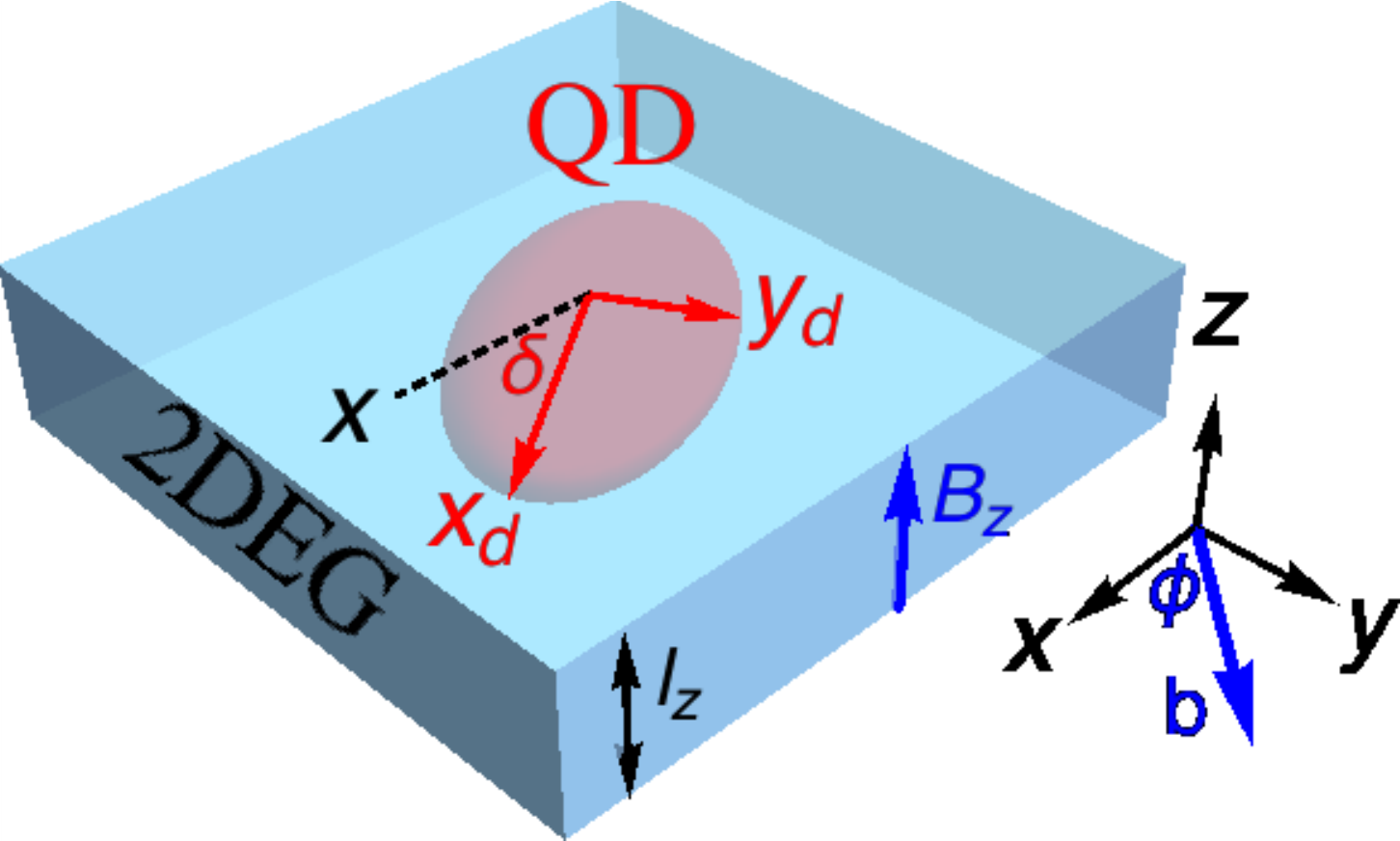} 
\caption{\label{fig:setup}
The schematic of the setup. The quantum dot (pink ellipsoid) is created by gates (not shown) in a 2DEG (blue slab). The 2DEG is a quasi-two-dimensional plane perpendicular to $\z \equiv [100]$ axis and has a nominal width $l_z$. The axes of the quantum dot potential, $\x_d$ and $\y_d$, are rotated by an angle $\delta $ with respect to the crystallographic axes $\x \equiv [100]$ and $\y \equiv [010]$. The magnetic field vector $\mathbf{B}$ has the out-of-plane component $\Bz$ and the in-plane component $\Bin$. The parameter $\phi$ denotes the angle of the vector $\Bin$ with the $\x$ axis.
}
\end{figure}

The basis functions in the three-dimensional space are defined as tensor product of the out-of-plane and in-plane terms, $| \alpha i \rangle \equiv |\alpha \rangle \otimes |i\rangle$. The corresponding energies are $E_{\alpha i} \equiv E_\alpha + E_i$. For further convenience, we define the aspect ratio $\eta$, as the ratio of the in-plane and subband energy spacings, $\eta={\rm min}\{E_x,E_y\}/E_z$. For dots embedded in 2DEGs, $\eta$ is a small parameter. The geometry is depicted in Fig.~\ref{fig:setup}.

\subsection{Unperturbed Zeeman energy}

\label{sec:uze}

\begin{figure}
\includegraphics[width=\columnwidth]{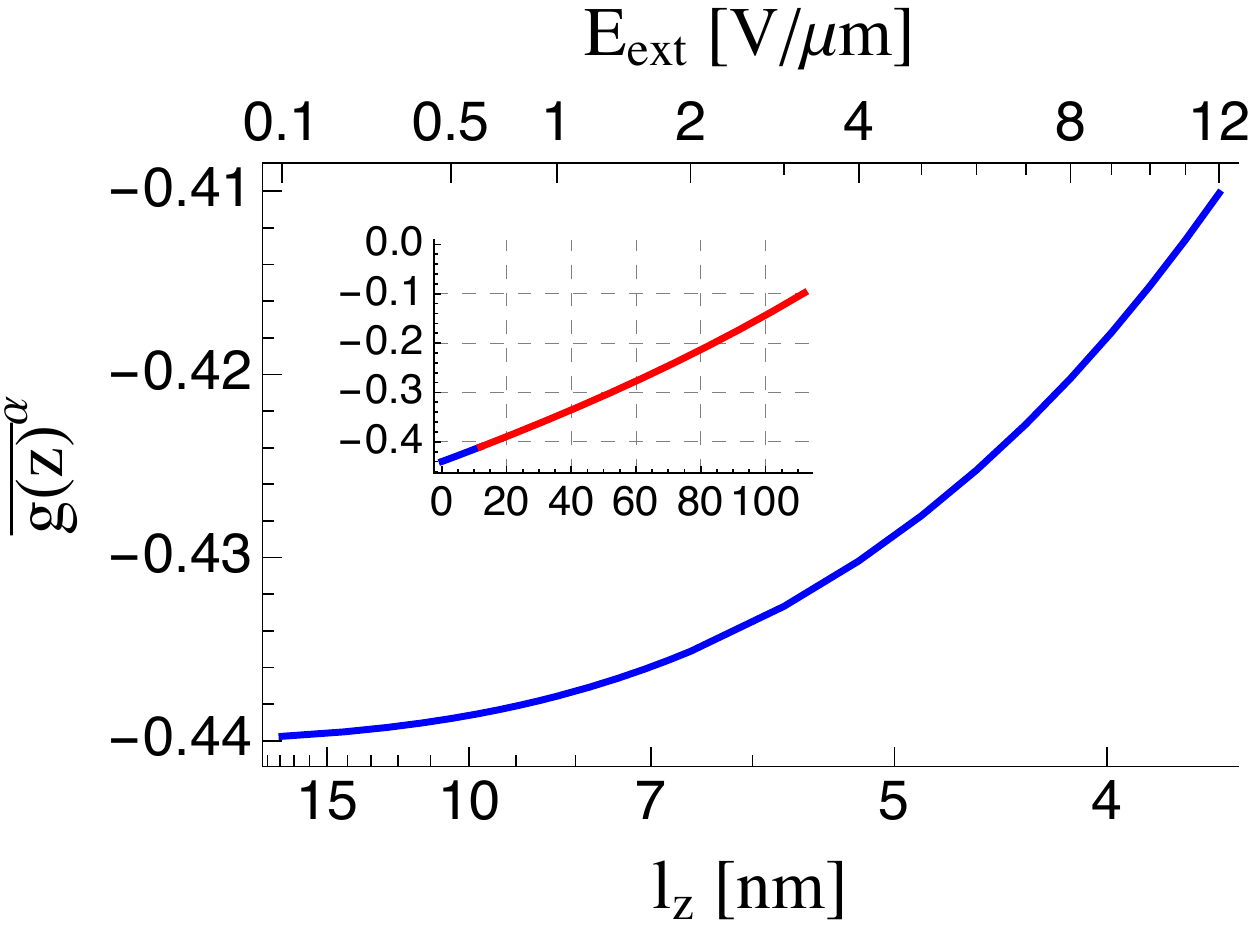}
\caption{\label{fig:g0} 
The lowest order approximation to the g-factor, showing the value of $\ave{g(z)}{\alpha}$ for the lowest subband, $\alpha=1$. The lower and upper x-axis shows, respectively, the nominal width of the 2DEG, and the interface electric field. They are related by $eE_{\rm ext} = \hbar^2/2ml_z^3$.
Inset: $\ave{g(z)}{\alpha}$ plotted for a larger range of the interface electric field $E_{\rm ext}$ (the inset x axis, given in V/$\mu$m). We show the electric field range for which at least one subband [a localized eigenstate of Eq.~\eqref{eq:Hz}] exists. The blue curve in the inset is the same as the blue curve of the main panel. We calculate the term numerically, solving for eigenstates of the triangular confinement potential with a finite conduction band offset $\delta E_c = E_{c}^{\rm A} - E_{c}^{\rm B}=300$ meV. We also use $m_{\rm A}=0.092\, m_e$, and $m_{\rm B}=0.067\, m_e$ with $m_e$ the free electron mass.
}
\end{figure}

At finite magnetic fields, the leading spin-dependent interaction in Eq.~\eqref{eq:H zeroth order} is the Zeeman term
\be
H_{\rm Z} = \frac{g(z)\mu_B}{2} {\bf B} \cdot \boldsymbol{\sigma},
\label{eq:HZ}
\ee
where the vector of Pauli matrices $\boldsymbol{\sigma}=(\sigma_x, \sigma_y, \sigma_z)$ is the electron spin operator, $\mu_B$ is the Bohr magneton, and 
the g-factor
\be
g(z) = \Theta(-z) g_A + \Theta(z) g_B,
\label{eq:gz}
\ee
is z-coordinate dependent, similarly to the effective mass. Taking the expectation of Eq.~\eqref{eq:HZ} in a chosen subband, 
the spin structure of the basis state $|\alpha i\rangle$ is described by
\be
H_Z^{(\alpha)} = \frac{\ave{g(z)}{\alpha}\mu_B}{2} {\bf B} \cdot \boldsymbol{\sigma}.
\label{eq:Zeeman0}
\ee
In Fig.~\ref{fig:g0}, we plot the g-factor averaged in the lowest subband as a function of the 2DEG width. Decreasing the width, the g-factor value departs from the bulk GaAs value towards the Al$_{x}$Ga$_{1-x}$As value, due to the penetration of the wave function into the barrier material. This effect is well known\cite{loss2000,jiang2001} and allows for an electrically tunable g-factor through designed material composition.\cite{salis2001,kato2003}

At this level of description, all states in a given subband have identical and isotropic g-factor. However, unlike for the mass dependence, we include the difference between exact and averaged interaction, 
\be
H_{\rm z} = H_Z - H_Z^{(\alpha)},
\label{eq:z}
\ee
among the perturbations considered below.

\subsection{Spin-independent perturbation}

The perturbation $H-H_0$  comprises the spin-independent and spin-dependent part. The first consists of the following terms 
\be
H^\prime_B = \frac{e}{m}\Ain \cdot \Pin + \frac{e^2}{2m} \Ain^2 \equiv H^\prime_1 + H^\prime_2,
\label{eq:Hmix}
\ee
which arise from the vector potential corresponding to the in-plane magnetic field, 
\be
\Ain = (z-z_0) (B_y, -B_x).
\label{eq:Ain}
\ee
Here, $z_0$ is a gauge choice, which will be specified later [below Eq. (33)].
In Ref.~\citenum{stano2017}, we have shown  how this perturbation affects the orbital structure of the quantum dot, that is, how the states $|\alpha i \rangle$ change. We found that the changes scale with the flux\footnote{We note that the name and interpretation of the quantity $\flux$ is motivated by the form of Eq.~\eqref{eq:phi}. No particular area $\lambda_z^2$ in the physical device can be pinpointed as defining this ``flux''.} due to the in-plane field
\be
\flux = \frac{e}{\hbar} \sqrt{B_x^2+B_y^2} \, \lambda_z^2,
\label{eq:phi}
\ee
serving as the small parameter of the perturbation theory. The flux depends on the effective 2DEG width,\cite{stern1968}
\be
\lambda_z^{4}= 2{\sum_{\beta \neq \alpha}} \frac{\hbar^2}{m} \frac{|z_{\alpha \beta}|^2}{E_\beta-E_\alpha},
\label{eq:lambdaz}
\ee
the behavior of which  was analyzed in detail in Ref.~\citenum{stano2017}.

\subsection{Spin-dependent perturbations}

In this article we are concerned with the effects of the in-plane magnetic field on the spin, rather than orbital, structure of the states. 
Such an analysis requires to expand the model by additional spin-dependent interactions. 
To this end, we adopt the Ogg-McCombe Hamiltonian,\cite{ogg1966, mccombe1969} which can be derived by the method of invariants. Namely, it comprises terms allowed by the $T_d$ symmetry group for the $\Gamma_6$ conduction band around its minimum at the wavevector $k=0$, up to the fourth order in the components of the kinetic momentum operator $\textbf{P}$.\cite{malcher1986} We use the coefficients of the invariant expansion terms that were obtained in the fourth order perturbation of the $k\cdot p$ theory including 14 bands\cite{braun1985} (counting also degeneracies and spin; if each at $k=0$ degenerate subspace is counted as one `level', the 14 band model is also called the 5L model\cite{pfeffer2006b}). 
This perturbative approach has been previously shown adequate in describing the conduction band g-factor in quantitative agreement with experiments.\cite{ivchenko1992,malinowski2000} 
We now list the spin dependent terms of the Ogg-McCombe Hamiltonian.

We first take the `standard' spin-orbit interactions. They comprise two terms, the bulk (Dresselhaus) term and the interface (Rashba) term. The former is
\be \begin{split}
H_{\rm D} = \frac{\gamma_c}{2\hbar^3} \Big( &\sigma_{x} \left\{ P_x, P_y^2 -P_z^2 \right\} + \sigma_{y} \left\{ P_y, P_z^2 - P_x^2 \right\} \\
 &+ \sigma_{z} \left\{ P_z, P_x^2 -P_y^2 \right\} \Big),
\end{split} 
\label{eq:HD}
\ee
where $\gamma_c$ is a material constant, and the curly brackets denote the anticommutator. The Rashba term is\footnote{In addition to this---``standard'' Rashba---term, a similar but higher-order-in-momenta term (proportional to the electric field and with cubic functions of momenta multiplying the Pauli matrices $\sigma_x$ and $\sigma_y$) is allowed by symmetry for electrons in GaAs. For holes, one can find an analogous term in literature under the name ``cubic Rashba'' term. Since it arises only in high order of the perturbation theory (for both holes and electrons in the fifth order in the extended Kane model\cite{winkler2003}), it is expected to be small. We do not consider such cubic term here.}
\be
H_{\rm R} =  \frac{\alpha_{R}(z)}{\hbar} \left(\sigma_y P_x - \sigma_x P_y \right),
\label{eq:HR}
\ee
where the pre-factor is well approximated by\cite{acta}
\be
\alpha_R(z) = \alpha_0 e E_{\rm ext}+\beta_{BA} \delta(z),
\label{eq:alphaRz}
\ee
with $\alpha_0$ and $\beta_{BA}$ being material constants, expressed through the bandstructure parameters by formulas given in Ref.~\citenum{stano2017} [see Eqs.~(C2) and (C6) therein].

The above spin-orbit terms are the leading spin-dependent corrections present at zero magnetic field. At finite magnetic fields, additional terms appear. As they do not have established names, we use the notation from Ref.~\citenum{braun1985}. The first term is isotropic in both spin and momentum separately,
\be 
H_{43} = \frac{e\gamma_{43}}{\hbar^3} \textbf{P}^2( {\bf B} \cdot \boldsymbol{\sigma} ).
\label{eq:H43}
\ee
The next two terms are anisotropic,
\be \begin{split}
H_{44} =  
\frac{e\gamma_{44}}{2\hbar^3}  & \Big[ \left( \left\{ P_x, P_y \right\} B_y + \left\{ P_x, P_z \right\} B_z \right) \sigma_x  \\
& +  \left( \left\{ P_y, P_z \right\} B_z + \left\{ P_y, P_x \right\} B_x \right) \sigma_y \\
& +  \left( \left\{ P_z, P_x \right\} B_x + \left\{ P_z, P_y \right\} B_y \right) \sigma_z \Big],
\end{split}
\label{eq:H44} 
\ee
and
\be 
H_{45} = \frac{e\gamma_{45}}{\hbar^3} \Big( P_x^2 B_x \sigma_x + P_y^2 B_y \sigma_y + P_z^2 B_z \sigma_z \Big).
\label{eq:H45}
\ee
In the above, $\gamma_{43}$, $\gamma_{44}$, and $\gamma_{45}$ are material dependent constants which are expressed through the bandstructure parameters in Refs.~\citenum{braun1985,lommer1986}. The terms in Eqs.~\eqref{eq:H43}--\eqref{eq:H45} have been essential to understand the dependence of the g-factor on the 2DEG width quantitatively, as well as to explain the anisotropy of the g-factor for magnetic fields in plane compared to magnetic fields out of plane.\cite{ivchenko1992, hannak1995, jeune1997, ivchenko1997, malinowski2000}
Whereas for the first effect, $H_{43}$ is the most important addition to the wave-function penetration effect shown in Fig.~\ref{fig:g0}, $H_{45}$ explains the in-plane versus the out-of-plane anisotropy upon noting that the heterostructure confinement makes the expectation value of the momentum operator components strongly different, $\langle P_z^2 \rangle \gg \langle P_{x,y}^2 \rangle$. The related light and heavy hole splitting by the confinement can be seen as the physical origin of this type of conduction g-factor anisotropy.\cite{ivchenko1992,kiselev1998} 

In the presence of both the electric and magnetic field, an additional term arises,\cite{winkler2003} which is not contained in the original Ogg-McCombe Hamiltonian. With an out-of-plane electric field and an in-plane magnetic field (the case to which we restrict ourselves below), this term is
\be 
H_{47} = \frac{e^2\gamma_{47} E(z)}{\hbar} (B_y \sigma_x + B_x \sigma_y),
\label{eq:H47}
\ee
where we use
\be 
E(z) =E_{\rm ext} + \frac{\beta_{BA}}{e\alpha_0} \delta(z),
\label{eq:Ez}
\ee
for the position dependent electric field, in analogy with Eq.~\eqref{eq:alphaRz}.

We use the following material parameters for $A=\textrm{Al$_{x}$Ga$_{1-x}$As}$ with $x=0.3$ and $B=\textrm{GaAs}$. The effective masses\cite{vurgaftman} $m_A=0.092 m_e$, $m_B= 0.067 m_e$, 
the g-factors $g_A=0.46$, $g_B=-0.44$,\cite{hannak1995} the spin-orbit strengths $\gamma_c=-10.6$ eV\AA$^3$,\cite{dettwiler2017} $\alpha_0=- 5.15$ \AA$^2$,\cite{knap1996} $\beta_{BA}=- 1.22$ eV\AA$^2$.\cite{acta} For the remaining coefficients we take\footnote{There seems to be an inconsistency or a typo in Refs.~\citenum{rossler1984, braun1985,lommer1986}. Namely, transforming $a_{43}$, $a_{44}$, and $a_{45}$ in Table 2 of Ref.~\citenum{lommer1986} into their dimensionful form, we get $\gamma_{43/4/5}$ as given here, in line with Ref.~\citenum{michal2017}. However, using Table 3 of Ref.~\citenum{braun1985} directly with the band parameters in Ref.~\citenum{rossler1984} we get $\gamma_{43}=1080$~eV\AA$^4$, $\gamma_{44}=-676.9$~eV\AA$^4$, and $\gamma_{45}=78.01$~eV\AA$^4$. We do not pursue the difference further, being of the order of one, which is not relevant for our purposes, and take the set with smaller values overall, as a conservative choice.}
$\gamma_{43}=493$~eV\AA$^4$, $\gamma_{44}=-433$~eV\AA$^4$, $\gamma_{45}=58$~eV\AA$^4$,\cite{lommer1986} and $\gamma_{47}=-5.2$~\AA$^3$.\cite{michal2017} 

Let us make the following comments for completeness. First, we do not include terms quartic in momenta in the Ogg-McCombe Hamiltonian (anharmonic and warping terms) as they do not directly couple to spin. They would change the basis (both the subbands and the in-plane orbital states), which would lead to minor renormalization of the numerical factors ($c$ and $\eta$ below). Second, we do not consider the z-dependence of the $k\cdot p$ coefficients $\gamma$ and take them as constants. On the one hand, these parameters do have different values in different materials, so that the penetration of the wave function into material A will renormalize them similarly to the g-factor and the effective mass. However, since the material values of these parameters have large uncertainties, their renormalization is of little practical consequence. Of interest here would be effects coming solely from their spatial dependence, which would be described by terms analogous to $H_{\rm z}$ in Eq.~\eqref{eq:z}. As we find below, the latter is negligible (it generates terms $g_{\rm z}$ in Fig.~\ref{fig:gs}), which a posteriori justifies taking $\gamma$'s as constant. Third, the one spin-orbit constant which we do not take constant in space is the Rashba coefficient. It is because it contains an explicit ``interface'' contribution, the $\delta$-function term in Eq.~\eqref{eq:alphaRz}. In principle, more interface contributions arise, corresponding to higher-order terms (in the electric field and in the momentum components) in the $k \cdot p$ theory. For example, Ref.~\citenum{alekseev2017} evaluates an additional interface term, similar in form to the Dresselhaus term. Nevertheless, the authors of that work find that unless the quantum well is very narrow, in GaAs the interface-Dresselhaus term is much smaller than the bulk-Dresselhaus term, justifying our choice again.\footnote{\label{fnt}The interface terms are important in silicon conduction band,\cite{golub2004} where the bulk spin-orbit coupling is very weak. See the introduction of Ref.~\citenum{alekseev2017} for an overview of the relevant literature on the interface spin-orbit terms.}

\subsection{The zeroth order spin-orbit interactions}

\label{sec:2E}

To simplify some formulas below, we denote the in-plane components of the magnetic field as $\Bin \equiv(B_x,B_y)$ and denote the angle that $\Bin$ makes with $\x$ as $\phi$.
Since our calculations rely on the expansion in the powers of the in-plane magnetic field, it is useful to introduce notation which explicitly reflects it. Namely, for the bulk Dresselhaus Hamiltonian, we denote as $H_{\rm d,n}$ the term proportional to $(\Binscalar)^n$. It can be calculated using the following recursive formula,
\begin{subequations}
\label{eq:powers}
\begin{eqnarray}
H_{\rm d,n} &= H_{\rm D}({\Bin=0}), \,\,\, \qquad &\textrm{if}\quad n=0,\\
H_{\rm d,n} &\,\,\,= \frac{1}{n}[\frac{e}{i \hbar} \Ain \cdot {\bf r},H_{\rm d,n-1}], \quad &\textrm{if}\quad n>0.
\end{eqnarray}
\end{subequations}
The highest non-zero term is with $n=3$. For the Rashba term, the same formulas can be used, though the terms beyond the linear one, $n=1$, are zero. The formula can be used also for the momentum dependent part of $H_{43,44,45}$, but we will not use such expressions explicitly. Rather, our main goal here is to connect to the standard notation for the spin-orbit terms without the orbital effects of the in-plane magnetic field. Namely, the lowest order spin-orbit interaction for subband $\alpha$ is obtained by taking the subband average of the $\Bin$-independent terms,
\be
H_{\rm d}^{(\alpha)} \equiv \ave{H_{\rm d, 0}}{\alpha}, \quad H_{\rm r}^{(\alpha)} \equiv \ave{H_{\rm r, 0}}{\alpha}.
\label{eq:Dresselhaus01}
\ee
In this way, we get the standard expressions of the linear-in-momenta, and the cubic-in-momenta terms,
\be \begin{split}
H_{\rm d}^{(\alpha)} =& \frac{\gamma_c}{\hbar^3} \ave{p_z^2}{\alpha} \Big( - \sigma_{x} p_{x} + \sigma_{y} p_{y} \Big)\\
&+\frac{\gamma_c}{2\hbar^3} \Big( \sigma_{x} \left\{ p_{x}, p_{y}^2 \right\} - \sigma_{y} \left\{ p_{y}, p_{x}^2 \right\} \Big),
\end{split}
\label{eq:Dresselhaus02}\ee
for the Dresselhaus term, and the linear-in-momentum terms,
\be
H_{\rm r}^{(\alpha)} = \frac{\ave{\alpha_R(z)}{\alpha}}{\hbar} \Big( \sigma_{y} p_{x} - \sigma_{x} p_{y} \Big),
\label{eq:Rashba0}
\ee
for the Rashba term.

\section{The perturbation theory}

We now explain our perturbation calculation. We aim at deriving spin-related corrections to the effective two-dimensional Hamiltonian for a given subband $\alpha$, reflecting the influence of the orbital effects of the in-plane magnetic field. To this end, we treat $H_0$, Eq.~\eqref{eq:H0}, as the unperturbed part, and the rest as the perturbation,
\be
H^\prime = H^\prime_B + H^\prime_S.
\ee
It comprises the spin independent part, Eq.~\eqref{eq:Hmix}, and  
\be
H^\prime_S= H_{\rm D} + H_{\rm R} + H_{\rm Z} + H_{43} + H_{44} + H_{45} + H_{47},
\label{eq:HS}
\ee
the spin dependent terms. 

Our results below list corrections which are linear in $H^\prime_S$, and up to the third order in the in-plane magnetic field $\Bin$.
However, to explain the essence of the approach, let us first consider a simplified case. Namely, up to the second order in the perturbation $H^\prime$, the matrix elements of the effective Hamiltonian for the $\alpha$-th subband are given by\cite{birpikus} 
\be
\begin{split}
H^{(\alpha)}_{i j} & =  \langle \alpha i | H^\prime | \alpha j\rangle +
 \frac{1}{2} {\sum_{\beta k}}^\prime
 \langle \alpha i | H^\prime | \beta k\rangle \\
&\times  \langle \beta k | H^\prime | \alpha j\rangle  \left( \frac{1}{E_{\alpha i}-E_{\beta k}} 
+ \frac{1}{E_{\alpha j}-E_{\beta k}} \right).
\end{split}
\label{eq:Lowdin}
\ee
The summation is over all $\beta$ and $k$ except for two pairs, $(\beta k) \neq (\alpha i)$, and $(\beta k) \neq (\alpha j)$. The formula is generalized to higher order and adjusted for our case in Appendix \ref{app:Lowdin}. For now, we look at terms arising in the simplified case described by the previous equation.

\subsection{Two examples of the effective spin-orbit interaction}

We now proceed with the evaluation of the effect on the spin beyond the lowest order term given in Eq.~\eqref{eq:Zeeman0}. We first present two examples, with which we motivate simplifications that we adopt in further steps to keep the results tractable. In both of these examples, we calculate the correction proportional to the first order of $H_{\rm d,1}$, so that it is linear in $\gamma_c$, and linear in $\Binscalar$. 
Let us first consider the $\beta=\alpha$ terms in Eq.~\eqref{eq:Lowdin} (we called such terms intra-subband in Ref.~\citenum{stano2017}). We get
\be
\begin{split}
H_{\rm d,1}^{(\alpha)} ({\rm intra}) 
&= \ave{H_{\rm d, 1}}{\alpha} - [\frac{e}{i \hbar} \ave{ \Ain  }{\alpha} \cdot {\bf r}, \ave{H_{\rm d, 0}}{\alpha}]\\
&=-\frac{e\gamma_c}{2\hbar^3} \left( B_{x} \sigma_{y} + B_{y} \sigma_{x} \right)  \\
& \times \left( \ave{\{ z-z_0, p_z^2 \} }{\alpha} - \{ \ave{z-z_0}{\alpha} , \ave{p_z^2}{\alpha} \}\right). 
\end{split}
\label{eq:example1}
\ee
This is the g-tensor correction derived in Ref.~\citenum{kalevich1993}. We make two simplifications based on this expression. First, we specify to the gauge $z_0=\ave{z}{\alpha}$ and denote $\Delta z= z-\ave{z}{\alpha}$. This choice makes the commutator term in Eq.~\eqref{eq:example1} zero, as well as analogous commutators in higher order terms, since their role is only to assure the gauge invariance of the result.\footnote{The choice $z_0=\ave{z}{\alpha}$ makes the subband-averaged vector potential zero, $\ave{\Ain}{\alpha}=0$. Should the general expressions be of interest, it is simplest to generate them by Taylor expanding the following identity $H\left(\Pin + e \ave{\Ain}{\alpha}\right) =  U^\dagger H(\Pin) U$, with $U=\exp\left( \frac{i}{\hbar}e \ave{\Ain}{\alpha}\cdot \Rin \right)$. We note in passing that our gauge choice is different from the one adopted in, for example, Refs.~\citenum{falko2005,miserev2017}.}
Second, we regroup the Pauli matrices into the following combinations,
\be
\begin{split}
B_{x} \sigma_{y} + B_{y} \sigma_{x}  &= \sin (2\phi) \boldsymbol{\sigma} \cdot \Bin
 - \cos (2\phi)   \boldsymbol{\sigma} \cdot \left( \Bin \times \z \right).
\end{split}
\label{eq:split}
\ee
The effective magnetic field defined by the second term is perpendicular to the effective magnetic field corresponding to the unperturbed Zeeman term, Eq.~\eqref{eq:Zeeman0}. As long as all the corrections are small (with respect to the unperturbed Zeeman energy), which is the case here, this off-diagonal term will only perturb the energy in the second order in its magnitude, which is beyond the perturbation order that we work in. The only consequence of the second term is a slight deflection (typically by less than $1^\circ$) of the quantization axis of the eigenstate spinor, an effect not of interest here.\footnote{The off-diagonal terms would be relevant for the electric-dipole spin-resonance, if the field generating them is periodically driven. 
We leave the analysis of electric manipulations for future work.} We therefore neglect below such off-diagonal terms. Equation \eqref{eq:example1} is then reduced to a contribution to the g-factor
\be
g_{\rm d,0}^{(\alpha)} ({\rm intra}) = -\frac{\lambda_d}{\lambda_z} c_1^{(\alpha)} \sin (2\phi).
\label{eq:example1result}
\ee
We parametrized the Dresselhaus constant by a length
\be
\lambda_d = \frac{4\gamma_c m_e}{\hbar^2},
\label{eq:lambdagamma}
\ee
with $m_e$ the electron mass in vacuum, and 
\be
c_1^{(\alpha)} = \frac{\lambda_{z}}{2 \hbar^2} \ave{\{ \Delta z, p_z^2 \} }{\alpha},
\label{eq:c1}
\ee
is a dimensionless factor. Both are plotted in Fig.~\ref{fig:cs}, and will be discussed below together with other terms of similar nature arising from other contributions.

Let us now take the $\beta\neq \alpha$ terms in Eq.~\eqref{eq:Lowdin}, called also inter-subband, corresponding again to the correction proportional to $H_{\rm d,1}$. We get 
\be
\begin{split}
H_{\rm d,1}^{(\alpha)} ({\rm inter}) &=  \frac{e\gamma_c}{m \hbar^3} {\sum_{\beta\neq \alpha}}  \frac{1}{E_{\alpha }-E_{\beta}} z_{\alpha\beta} \ave{ p_z^2 }{\beta\alpha} \\
& \quad \times \left\{  \left( B_y p_x - B_x p_y \right) ,\left( -\sigma_x p_x + \sigma_y p_y \right) \right\}.
\end{split}
\label{eq:example2}
\ee
Unlike in Eq.~\eqref{eq:example1}, the effective Hamiltonian now contains both spin and in-plane momentum operators, and is thus an effective spin-orbit interaction. It inherits the angular anisotropies from the original spin-orbit interactions, as well as the reference to the direction of the magnetic field. One should therefore expect anisotropies in, for example, the corresponding spin relaxation rates,\cite{scarlino2014} or the electric-dipole spin resonance amplitudes,\cite{golovach2006,stano2008} which are different to the anisotropies corresponding to the zeroth order spin-orbit fields. Even though the detailed analysis is beyond the scope here, we expect that this effect is minor. Namely, the most important attribute of these higher order `spin-orbit' interactions is that, being generated by a magnetic field, they are not time-reversal symmetric. Unlike the zeroth order ones, they can therefore contribute to the g-factor in the lowest order, as we have seen already for Eq.~\eqref{eq:example1}. We therefore restrict ourselves to evaluating only this leading-order correction to the energy, by taking the expectation value in the unperturbed orbital eigenstate $i$ of the subband $\alpha$. Equation \eqref{eq:example2} then reduces to a g-factor correction [taking again only the diagonal part, similar to the first term in Eq.\eqref{eq:split}]
\be
g_{\rm d,0}^{(\alpha,i)} ({\rm inter}) = \frac{\lambda_d}{\lambda_z} c_2^{(\alpha)} \left(-\eta_+^{(i)} \sin (2\phi) + \eta_-^{(i)} \sin(2\delta) \right),
\label{eq:gd1inter}
\ee
with $c_2$ another dimensionless constant (all these constants are listed in Appendix \ref{app:cs}).
The presence of the momentum operators in Eq.~\eqref{eq:example2} makes this correction, unlike the one in Eq.~\eqref{eq:example1result}, dependent on the in-plane size and orientation of the dot, through
\be
\eta_{\pm}^{(i)} =\frac{\lambda_z^2}{2\hbar^2} \langle  i |(\Pin \cdot \x_d)^2  \pm (\Pin \cdot \y_d)^2 | i \rangle.
\label{eq:etas}
\ee
Quite naturally, the part which does not refer to the dot orientation [the first term in the bracket in Eq.~\eqref{eq:gd1inter}] is proportional to the quantity characterizing the average size, $\eta_+$, while the part which refers to the dot orientation [the second term in the bracket in Eq.~\eqref{eq:gd1inter}] is proportional to the orbital asymmetry, $\eta_-$, of state $i$. For illustration, assuming zero out-of-plane magnetic field, and neglecting here the small effects of the effective mass renormalization,\cite{stano2017} these two parameters become
\be
\eta_{\pm}^{(i)} =\frac{\lambda_z^{2}}{2l_x^{2}} \left( n_x^{(i)}+\frac{1}{2}\right) \pm \frac{\lambda_z^2}{2l_y^{2}} \left(n_y^{(i)}+\frac{1}{2}\right),
\ee
where the quantum numbers correspond to the state $i$, as defined below Eq.~\eqref{eq:Pin}. Specifying further to the ground state, we got
\be
\eta_{\pm}^{(\rm ground)} =\lambda_z^{2} (l_x^{-2}\pm l_y^{-2})/4.
\ee
For a dot which is circularly symmetric in the 2DEG plane, the basis can be always chosen such that $\eta_-^{(i)}=0$ for all $i$. For a general dot, these two factors fulfill $\eta_- \lesssim \eta_+ \propto \eta$, so that they are small, of the order of the aspect ratio.

\subsection{What is calculated: corrections to the g-factor}

Based on the two presented examples, we now set our goals for the calculations, organization of the results, and their analysis. We aim at the corrections to the g-factor\footnote{The g-factor will be a function of the magnetic field and we understand it here as the ratio of the Zeeman energy and the magnetic field. Its value at $B=0$ is to be understood as measured in the limit $B\to 0$, rather than directly at $B=0$.} for a chosen subband $\alpha$ and orbital state $i$, obtained as the expectation value of the effective spin-orbit interaction generated by the in-plane field in this specific state.  We restrict ourselves to the lowest order in $H_S^\prime$, or, loosely denominating the prefactors in various terms of $H_S^\prime$ with a common name, in spin-orbit couplings. We choose the simplest gauge, $z_0=\ave{z}{\alpha}$, and assume zero out-of-plane magnetic field for simplicity. Finally, we calculate the corrections up to the third order in the in-plane field, which is the highest order of the magnetic field appearing in $H^\prime_S$, Eq.~\eqref{eq:HS}. Note that it requires to include also the third order perturbation terms, going beyond Eq.~\eqref{eq:Lowdin}, as explained in Appendix \ref{app:spin-dep}.

Proceeding in this way, we are therefore neglecting terms being higher order in spin-orbit interactions (we estimate the largest such in Appendix \ref{app:lin-lin} and show that they are very small), terms of higher than the third order in the magnetic field, and terms admixing different in-plane orbitals. We calculate also the off-diagonal g-tensor components, but give them only in Appendix \ref{app:off-diagonal}. 
In the derivations, we neglect the in-plane with respect to the subband excitation energies, which brings in the derived formulas an error of the order of $\eta$, the aspect ratio.

In the derived expressions, we are interested in several aspects. The most important question is, how large corrections to the g-factor should one expect upon applying an in-plane field. However, the simple magnitude comparison is not all, as the arising terms differ qualitatively in the dependence on: the magnetic field magnitude (either constant or growing quadratically with the in-plane field), the 2DEG width (both increase and decrease with $\lambda_z$ are possible), the heterostructure symmetry (several terms do not arise in a symmetric quantum well), and the magnetic field orientation (the terms are either isotropic, or anisotropic but independent on the dot orientation, or dependent on it).

\section{Results for \lowercase{g}-factor corrections}

\label{sec:results}

We now list the obtained results. We first list the individual corrections, originating in the respective terms of the spin-orbit Hamiltonian. After that, we comment on the components of the corrections, which shed light on the overall scales and tendencies. Finally, we present the total g-factor correction, a sum of all contributions.

\mysubsection{Individual corrections}

Here, we list the individual terms. The left hand side of each equation gives the g-factor correction $g_{x,n}$ where $x$ denotes the origin of the term, with $x={\rm d}$ for the Dresselhaus term, $x={\rm r}$ for the Rashba term, 
$x=43-47$ for the corresponding $H_x$, and $x={\rm z}$ for $H_{\rm z}$.
The integer $n$ denotes the power of the magnetic field on the right hand side. To simplify the notation, we omit the subband index $\alpha$ and orbital state index $i$. On the right hand side, the subband dependence enters through the dimensionless factors $c$, such as the one in Eq.~\eqref{eq:c1}, and the orbital-state dependence enters through the factors $\eta^{(i)}_\pm$, defined in Eq.~\eqref{eq:etas}.

The contributions from the Dresselhaus interaction are
\begin{subequations}
\label{eq:gd}
\begin{eqnarray}
g_{\rm d,0}&=& \frac{\lambda_d}{\lambda_z} \Big( (-c_1 - c_2 \eta_+)\sin (2\phi) +c_2 \eta_- \sin (2\delta) \Big), \phantom{xxxxxx}\label{eq:gd0}\\
g_{\rm d,2}&=& \frac{\lambda_d}{\lambda_z} \Phi^2  \Big( [c_3-c_5  + (3 c_4-c_6+c_{14}) \eta_+] \sin (2\phi)\nonumber\\
&&  \,\,\, -(3 c_4 + c_{14}) \eta_- \sin (2\delta) \nonumber\\ 
&&  \,\,\, -c_4 \eta_-  \cos (2\phi) \sin(2\phi-2\delta) \nonumber\\ 
&& \,\,\, +c_6 \eta_- \sin (2\phi) \cos(2\phi-2\delta) \nonumber\\ 
&& \,\,\, + c_1 c_{16} \sin(2\phi) [\eta_+ - \eta_- \cos (2\phi-2\delta)]  \Big).
\label{eq:gd2}
\end{eqnarray}
\end{subequations}
The contributions from the Rashba interaction are
\begin{subequations}
\label{eq:gr}
\begin{eqnarray}
g_{\rm r,0}&=& -\xi_r  \Big( c_{10} - 4c_{11} [\eta_+ - \eta_- \cos (2\phi-2\delta)] \Big),\phantom{xxxxxxxx}
\label{eq:gr0}\\
g_{\rm r,2}&=& \frac{\lambda_z}{\lambda_r} \Phi^2  \Big( c_4 + c_7  [\eta_+ - \eta_- \cos (2\phi-2\delta)] \Big) \nonumber \\
&& -\xi_r \Phi^2 \Big(c_{12} + (3 c_{13} + c_{15} -c_{10} c_{16}) \nonumber \\
&& \hspace{0.8in} \times [\eta_+ - \eta_- \cos (2\phi-2\delta)] \Big). \label{eq:gr2}
\end{eqnarray}
\end{subequations}
The contributions from $H_{43}$ are
\begin{subequations}
\label{eq:g43}
\begin{eqnarray}
g_{43,0}&=& \frac{\lambda_{43}^2}{\lambda_z^2} \Big( c_{17} + 2 \eta_{+} \Big), \label{eq:g430}\\
g_{43,2}&=& \frac{\lambda_{43}^2}{\lambda_z^2} \Phi^2 \Big( c_{18} + c_{20} + (4 c_{19} + 3c_{21} -c_{17} c_{16} ) \nonumber \\
&& \hspace{0.5in} \times [\eta_+  - \eta_- \cos (2\phi-2\delta)] \Big). \label{eq:g432}
\end{eqnarray}
\end{subequations}
The contributions from $H_{44}$ are
\begin{subequations}
\label{eq:g44}
\begin{eqnarray}
g_{44,0}&=& \frac{\lambda_{44}^2}{\lambda_z^2} \eta_- \sin(2\phi) \sin (2\delta) , \label{eq:g440} \\
g_{44,2}&=& -\frac{\lambda_{44}^2}{\lambda_z^2} \Phi^2 \Big( \frac{c_{18}}{4} [1  - \cos (4\phi)] + 2c_{19} \sin(2\phi) \nonumber \\
&& \hspace{0.5in} \times  [\eta_+ \sin(2\phi) - \eta_- \sin (2\delta)] \Big). \label{eq:g442}
\end{eqnarray}
\end{subequations}
The contributions from $H_{45}$ are
\begin{subequations}
\label{eq:g45}
\begin{eqnarray}
g_{45,0}&=&  \frac{\lambda_{45}^2}{\lambda_z^2} [\eta_+  - \eta_- \cos (2\phi) \cos (2\delta)]  ,\label{eq:g450}\\
g_{45,2}&=& \frac{\lambda_{45}^2}{\lambda_z^2}  \Phi^2 \Big( \frac{c_{18}}{4} [1  - \cos (4\phi)] + 2c_{19} \sin(2\phi) \nonumber \\
&& \hspace{0.5in} \times  [\eta_+ \sin(2\phi) - \eta_- \sin (2\delta)] \Big). \label{eq:g452}
\end{eqnarray}
\end{subequations} 
The contributions from $H_{47}$ are
\begin{subequations}
\label{eq:g47}
\begin{eqnarray}
g_{47,0} &=& \Big( \frac{\lambda_{47}^3}{l_{z}^3} - \frac{\lambda_{47}^{\prime}}{\lambda_{z}} c_{22} \Big) \sin(2\phi), \label{eq:g470}\\
g_{47,2} &=&  -\frac{\lambda_{47}^{\prime}}{\lambda_{z}}  \Phi^2  \sin(2\phi)
 \Big(  c_{23} + (3 c_{24} -c_{22} c_{16})  \nonumber \\  && \hspace{0.7in}  \times [\eta_+  - \eta_- \cos (2\phi-2\delta)] \Big).
\label{eq:g472}
\end{eqnarray}
\end{subequations}
Finally, the bulk g-factor inhomogeneity gives
\begin{subequations}
\label{eq:gi}
\begin{eqnarray}
g_{\rm z,0}&=& 0,
\label{eq:gi0}\\
g_{\rm z,2}&=& \Phi^2 \Big( c_8  + c_{9} [\eta_+  - \eta_- \cos (2\phi-2\delta)] \Big). \phantom{xxxxxx}
\label{eq:gi2}
\end{eqnarray}
\end{subequations}
For completeness, we also define the ``penetration'' correction,
\be
g_{\rm p} = \ave{g(z)}{\alpha=1} - g_B,
\label{eq:gz00}
\ee
for the deviation of the lowest-subband-averaged g-factor from the bulk value in material B (GaAs) due to the leakage of the wavefunction into material A. 
Rather than giving a formula, we calculate it 
numerically. It was already explained in Sec.~\ref{sec:uze} and plotted in Fig.~\ref{fig:g0}: it is neither magnetic-field magnitude nor direction dependent.

\mysubsection{Correction components}

\begin{figure}
\includegraphics[width=\columnwidth]{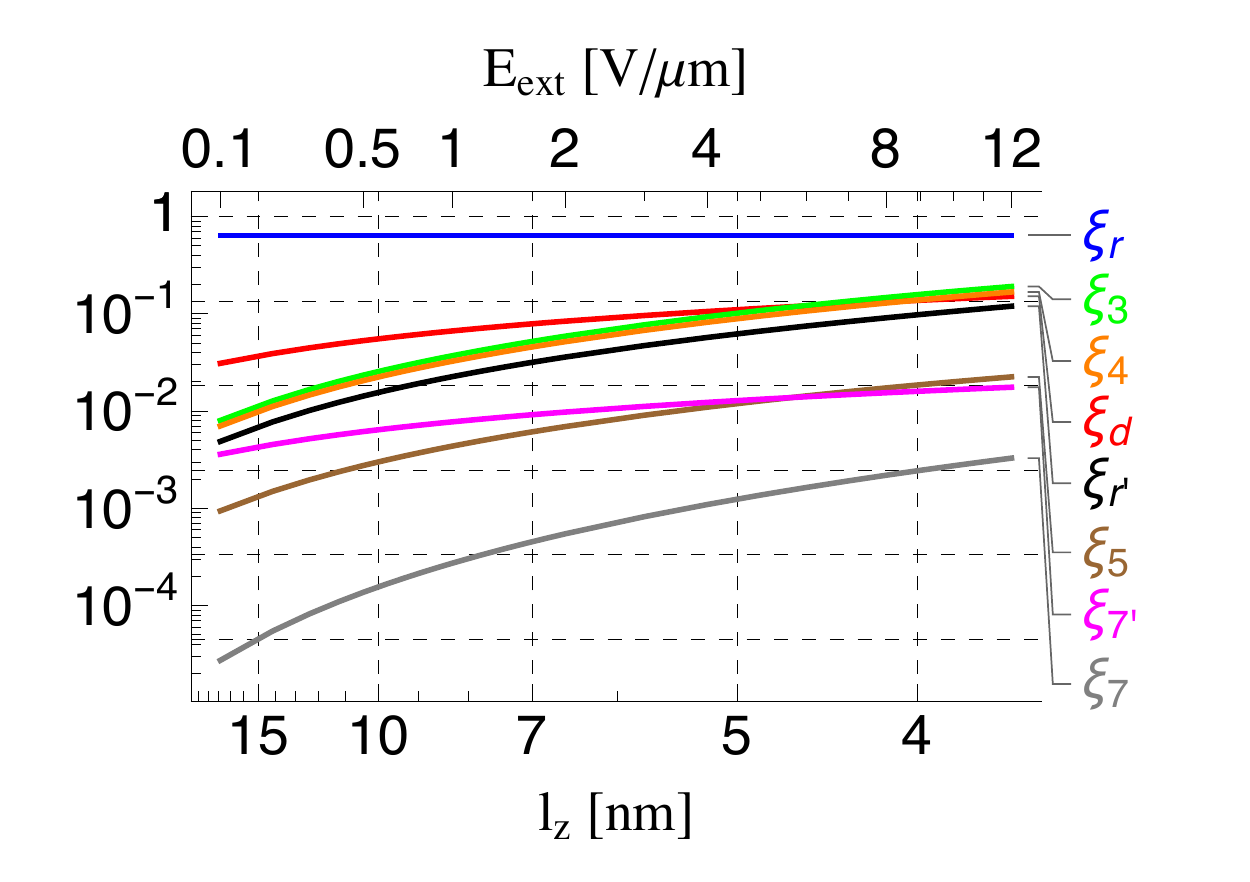}
\caption{\label{fig:xis} 
The scale parameters, defined in Tab.~\ref{tab:lambdas}, as a function of the 2DEG width.}
\end{figure}

\begin{figure}
\includegraphics[width=\columnwidth]{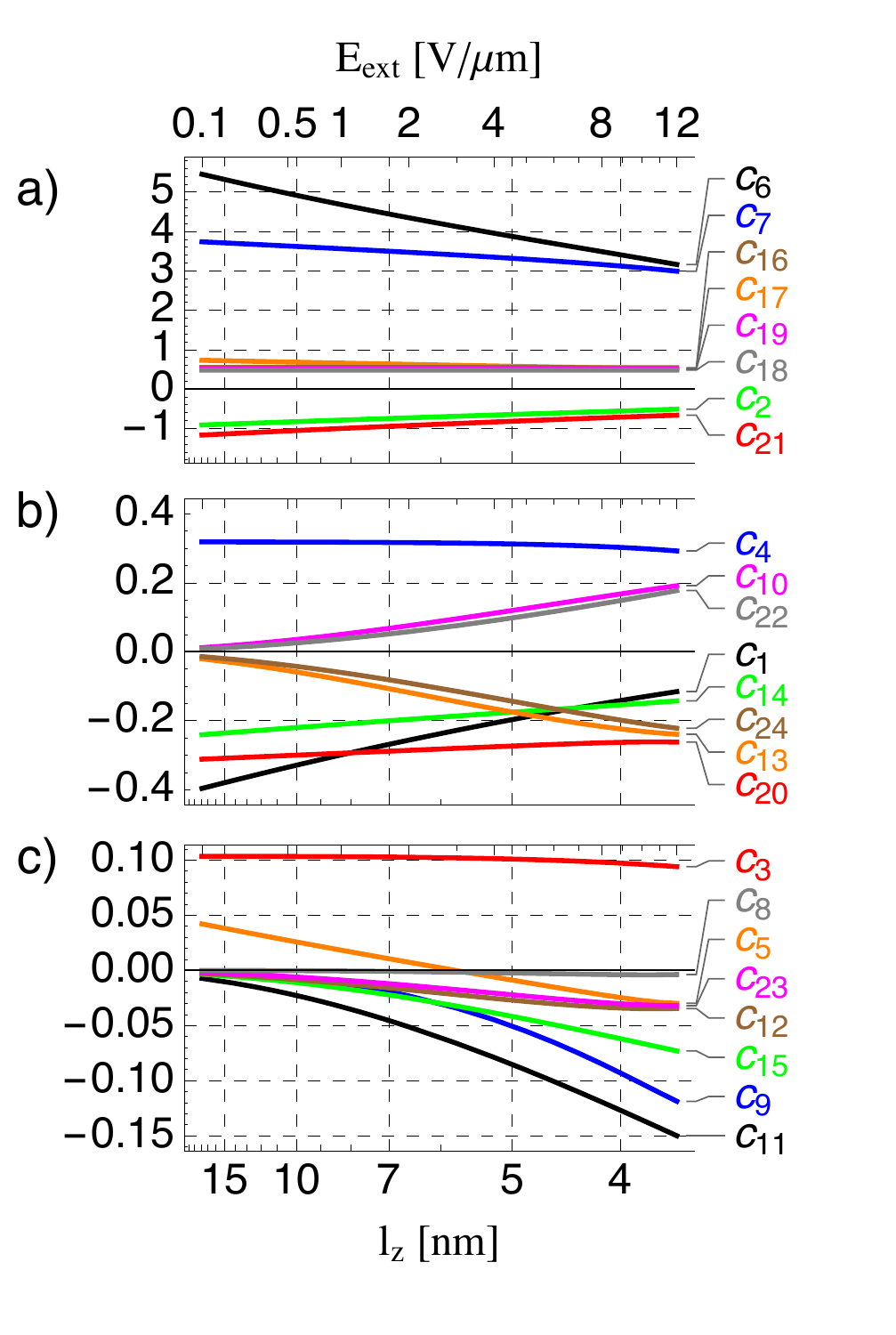}
\caption{\label{fig:cs} 
The dimensionless constants $c_n$ in the lowest subband $\alpha=1$.
We find that for $c$'s, the most natural parameter is the nominal width $l_z$, which is the parameter used for the lower x axis. The corresponding interface electric field is the upper x axis. The two are related by Eq.~\eqref{eq:lz}.
 }
\end{figure}

Let us first make some general comments on the above formulas. They split a g-factor correction to several dimensionless constituents, namely the strength, the magnetic field dependence, the numerical factors $c$ and $\eta$, and the angular dependence. The strengths can be expressed as a certain power of the ratio of a length characteristic for each interaction, and, essentially, the 2DEG width.\footnote{There is one exception: $\xi_r$ is already dimensionless, so it is not useful to recast it as a length scale. Also, the scale dividing $\lambda_{47}$ is $l_z$ rather than $\lambda_z$. However, for the triangular confinement the difference between the latter two is completely negligible (see Tab.~I in Ref.~\citenum{stano2017}).}
The lengths are summarized in Tab.~\ref{tab:lambdas}, and the corresponding strengths are plotted in 
Fig.~\ref{fig:xis}.
Concerning the magnetic field dependence, we obtained terms which are either constant, or grow quadratically with the in-plane flux. There are no terms linear in the magnetic field.\footnote{\label{fnt:B2}Such linear-in-B terms were reported in theory\cite{lommer1985,falko2005} and experiments.\cite{dobers1988,heberle1994,hanson2003,nefyodov2011b} In all cases where the origin can be identified, it corresponds to the limit of Landau levels, meaning that the orbital effects of the magnetic field dominate the electrostatic confinement (either within the 2DEG plane,\cite{lommer1985} or even perpendicular to it\cite{falko2005}). Such a limit corresponds to the magnetic field having beyond-perturbative influence on the excitation energies (whether in-plane or subband ones); that is, some of the excitation energies become linear in the field, $E_{\alpha i}-E_{\beta j} \sim |B|$. To put it in another way, there are no linear-in-$B$ terms in the g-factor as long as the basis in which the pertubative calculation is done is time-reversal symmetric, which is the case here.}
The constants $c$ are not expected to display any systematic dependence, given the differences in their origins (see Appendix \ref{app:cs} for explicit expressions). They are plotted on Fig.~\ref{fig:cs}. The factors $\eta_\pm$ give useful information about relative scales: for a nearly symmetrical dot, one can assume the hierarchy $1 \gg \eta_+ \gg | \eta_- | $. Finally, let us note the angular anisotropy. The terms which have cylindrical symmetry in the bulk, that is $x=\textrm{r},\, 43$, and $\textrm{z}$, result in corrections which are either isotropic, or anisotropic only due to the shape of the quantum dot. The latter terms depend on the relative orientation of the magnetic field with respect to the dot potential soft axis, through a common factor
\be
\eta_+ - \eta_- \cos(2\phi-2\delta).
\ee
The remaining terms, which do not have cylindrical symmetry in the bulk, contain different factors. They relate separately to the crystallographic axes, or the quantum dot axes. We expect that these properties remain valid in higher orders of the perturbation theory.

\begin{table}
\begin{tabular}{ccccccc}
\hline \hline
origin& definition & unit & d-full& scale&d-less&$|\lambda | [$\AA$]$\\
\hline
Dress.&  $ {4\gamma_c m_e} / {\hbar^2}$  & length$^1$ &$\lambda_d$ & $\lambda_z$&$\xi_d$&5.5\\
Rashba &  ${ 4\alpha_0 e E_{\rm ext} m_e }/{ \hbar^2 }$ & length$^{-1}$ &$\lambda_r^{-1}$ & $\lambda_z^{-1}$&$\xi_{r^\prime}$&1730\\
Rashba & $ { 4\beta_{BA} m_e }/{ \hbar^2 }$ & length$^0$ &$\xi_r$  & 1&$\xi_r$&-\\
43 &  $4{\gamma_{43} m_{e}  }/{ \hbar^2 }$ & length$^2$ &$\lambda_{43}^2$ & $\lambda_z^2$&$\xi_{3}$&16\\
44 &  $4{\gamma_{44} m_{e}  }/{ \hbar^2 }$ & length$^2$ &$\lambda_{44}^2$ & $\lambda_z^2$&$\xi_{4}$&15\\
45 &  $4{\gamma_{45} m_{e}  }/{ \hbar^2 }$ & length$^2$ &$\lambda_{45}^2$ & $\lambda_z^2$&$\xi_{5}$&5.5\\
47&   $2{\gamma_{47}  m_e}/{m_B}$ & length$^3$ &$\lambda_{47}^3$ & $l_z^3$&$\xi_{7}$&5.3\\
47&  ${ \xi_r \gamma_{47} }/{ \alpha_0 }$ & length$^1$ &$\lambda_{47}^{\prime}$ & $\lambda_z$&$\xi_{7^\prime}$&0.65\\
\hline \hline
\end{tabular}
\caption{\label{tab:lambdas}
Material constants parameterizing the g-factor corrections. Every $\lambda$ has the dimension of length, every $\xi$ is dimensionless. Column 1 gives the terms origin, column 2 its definition using $k \cdot p$ parameters and column 3 the unit of the expression in column 2. The units show that each of the dimensionfull parameters given in column 2 can be expressed as a length raised to some integer power. Such lengths are defined in column 4. They enter the g-factor corrections in a dimensionless form denoted by column 6, which is equal to the dimensionfull expression divided by the scale given in column 5. Finally, the last column gives the absolute value of the scale $\lambda$ introduced in column 4. We exemplify these definitions taking the term ``43'': $\xi_3=\lambda_{43}^2 / \lambda_z^2=4{\gamma_{43} m_{e}  }/{ \hbar^2 \lambda_z^2}$, and $|\lambda_{43}|\approx 16$\AA. The length $\lambda_r$ (row 2) depends on the interface electric field and the value $\lambda_{r}=173$ nm is for $E_{\rm ext}=2.14$ V/$\mu$m. Finally, the parameter $\xi_r$ (row 3) is dimensionless without introducing any scale and therefore 1 is used for the latter.
}
\end{table}

\mysubsection{Correction hierarchy}

We now turn to quantitative analysis. First, from Table \ref{tab:lambdas} and, more directly, from Fig.~\ref{fig:xis}
, one can see that the correction strengths generally grow upon narrowing the 2DEG (the only exception is $\xi_r$ which remains constant), but with different slopes. On the other hand, the magnetic flux also diminishes as the 2DEG is made narrower. 
Finally, the dependence gets further involved due to a non-systematic behavior, and a wide scale variation, of the dimensionless constants $c$, see Fig.~\ref{fig:cs}.
 Therefore, to nail down the importance hierarchy of the terms, it is easiest to look directly at the full terms, plotted in Fig.~\ref{fig:gs}. 
 
Figure \ref{fig:gs}(a)-(d) reveals the relative importance of the terms for a wide range of magnetic fields and 2DEG widths. We conclude that, concerning the g-factor corrections for the considered range of the interface fields, it is enough to include the Rashba, Dresselhaus, and $H_{43}$ terms. For very narrow 2DEG, the effect of the wave-function penetration into the barrier might be sizable, but it does not have to be considered beyond the averaging performed in Eq.~\eqref{eq:Zeeman0} and reflected in Fig.~\ref{fig:g0}. Similarly, the influence of $H_{45}$, $H_{47}$, and, perhaps with an exception of very wide 2DEGs and high magnetic fields, also $H_{44}$, is negligible. Focusing on the relevant terms [close-ups are shown in Fig.~\ref{fig:gs}(b) and (d)], Rashba and $H_{43}$ dominate the Dresselhaus term the narrower the 2DEG becomes. Their magnetic field dependence is rather weak if the flux is small, $\Phi \ll 1$, as expected. Once the flux becomes of order one, the magnetic-field dependence is more pronounced, and the interference of the field-independent and the field-dependent terms can lead to sign reversals, exemplified as sharp dips visible on Fig.~\ref{fig:gs}(a). 

The angular dependence is shown on Fig.~\ref{fig:gs}(e)-(f). The variation is dominated by the Dresselhaus term---even though this term is not largest in magnitude---with extrema related to the crystallographic axes (along [110] and [1$\overline1$0]). The same behavior was established for 2DEG in theory\cite{kalevich1993} and experiments.\cite{eldridge2011,nefyodov2011a,nefyodov2011b} The variation of the other two terms are much smaller, roughly by the factor $\eta_-$, with extrema related to the dot potential axes. As a result, the g-factor of a quantum dot should show a sizable directional dependence, with a minimum along [1$\overline{1}$0] (assuming the sign of $\gamma_c$ is negative).\footnote{The position of the minimum will be slightly shifted away from [110] by other terms that have extrema along different directions. As these additional variations are much smaller, the shift will be accordingly small, see Fig.~\ref{fig:gtot}.} In panel (e), the magnitude of the predicted directional variation is more than 10\% of the full g-factor value.
Since the overall angular variation mainly arises from the bulk-Dresselhaus term, its relative importance increases with the 2DEG width. Indeed, in Fig.~\ref{fig:gs}(f) the angular dependence for a wider 2DEG indicates that the directional variation becomes comparable to the corrections due to the Rashba and $H_{43}$ terms.

Next, we note the overall sign, looking at Fig.~\ref{fig:gs}(e). Assuming that both $-\alpha_0$ and $\gamma_{43}$ are indeed positive, both related corrections are positive, and diminish the magnitude of the negative g-factor in bulk GaAs. This is indeed the typical case seen in experiments. For the values assumed in panel (e), which correspond roughly to the interface parameters deduced from the experiment in Ref.~\citenum{camenzind2}, we would get the average g-factor of around $-0.33$. In that experiment, $|g| \approx 0.36$ was fitted from spectral data.

\mysubsection{The sum of all contributions}

We summarize the predictions of our formulas plotting the total g-factor [including the sum of all contributions in Eqs.~\eqref{eq:gd}--\eqref{eq:gz00}] in Fig.~\ref{fig:gtot}. For the parameters typical for the experiments in Refs.~\citenum{camenzind1,camenzind2}, the g-factor as a function of the magnetic field direction looks as in Fig.~\ref{fig:gtot}(a). The curve is characterized by three numbers, the average value, the magnitude of the variation, and the position of the maximum. We plot these quantities as functions of the 2DEG width and magnetic field in Fig.~\ref{fig:gtot}(c)-(f). Fig.~\ref{fig:gtot}(c) shows that the correction grows upon narrowing the 2DEG, in line with the behavior seen in Fig.~\ref{fig:gs}(a). On the other hand, the variation magnitude is non-monotonic. It is biggest at around $l_z\approx 7$ nm, where it is almost half of the total correction. The position of the maximum deviates only slightly from $[1\overline{1}0]$, in line with Fig.~\ref{fig:gs}(e). Panel (e) shows the effects of the in-plane magnetic field magnitude. Comparing the solid and dashed curves, one can confirm that the g-factor variations upon changing the magnetic field are bigger in wider 2DEG. 

Finally, we notice that the g-factor corrections are more pronounced for excited states: the variation can become larger than the average correction sooner, while the position of the extrema can shift by larger angles. These facts are illustrated in Fig.~\ref{fig:gtot}(d),(f).
Figure \ref{fig:gtot}(f) shows the largest corrections are due to the same terms as for the ground state. However, the actual values of these contributions are appreciably different in their magnitude. One can therefore expect substantial differences in the g-factor renormalizations for different states of the quantum dot.

\begin{widetext}

\begin{figure}[h!]
\includegraphics[width=0.49\columnwidth]{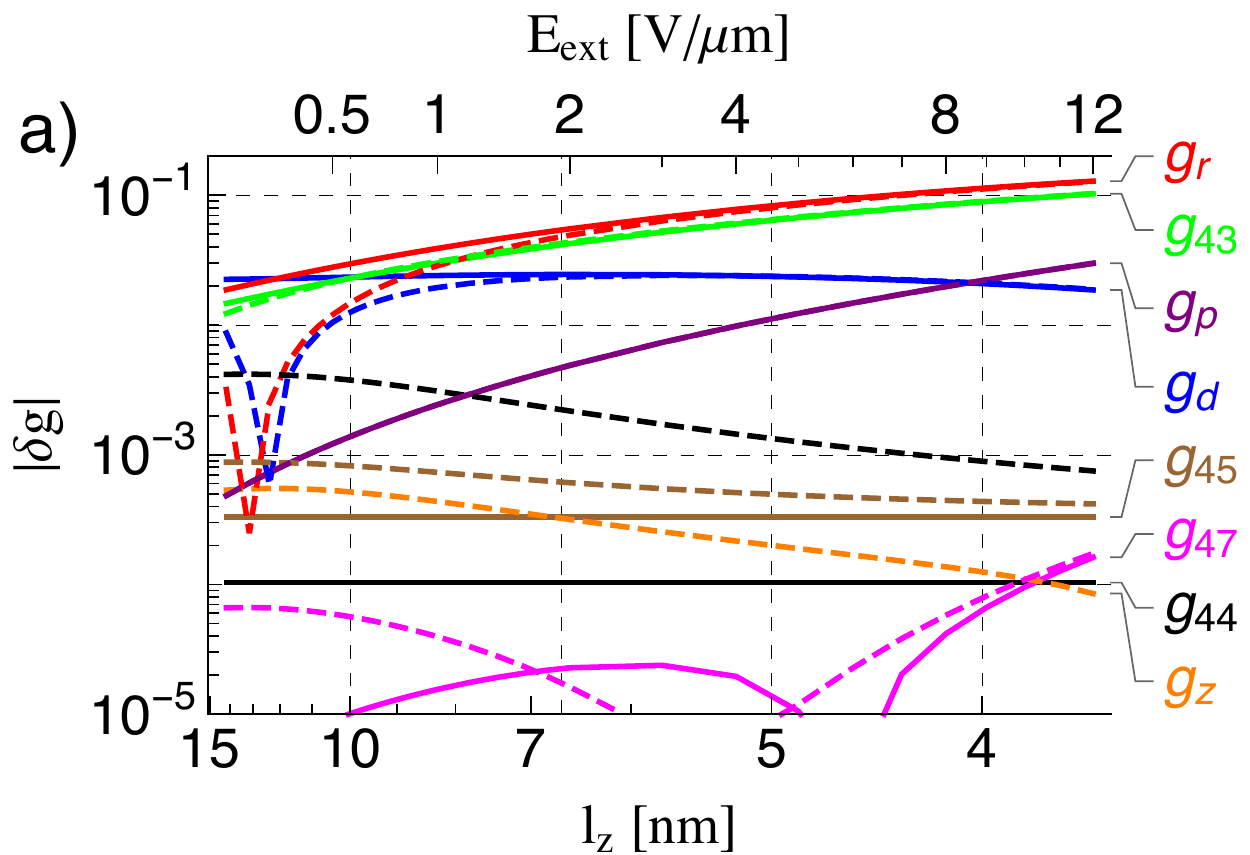}\includegraphics[width=0.49\columnwidth]{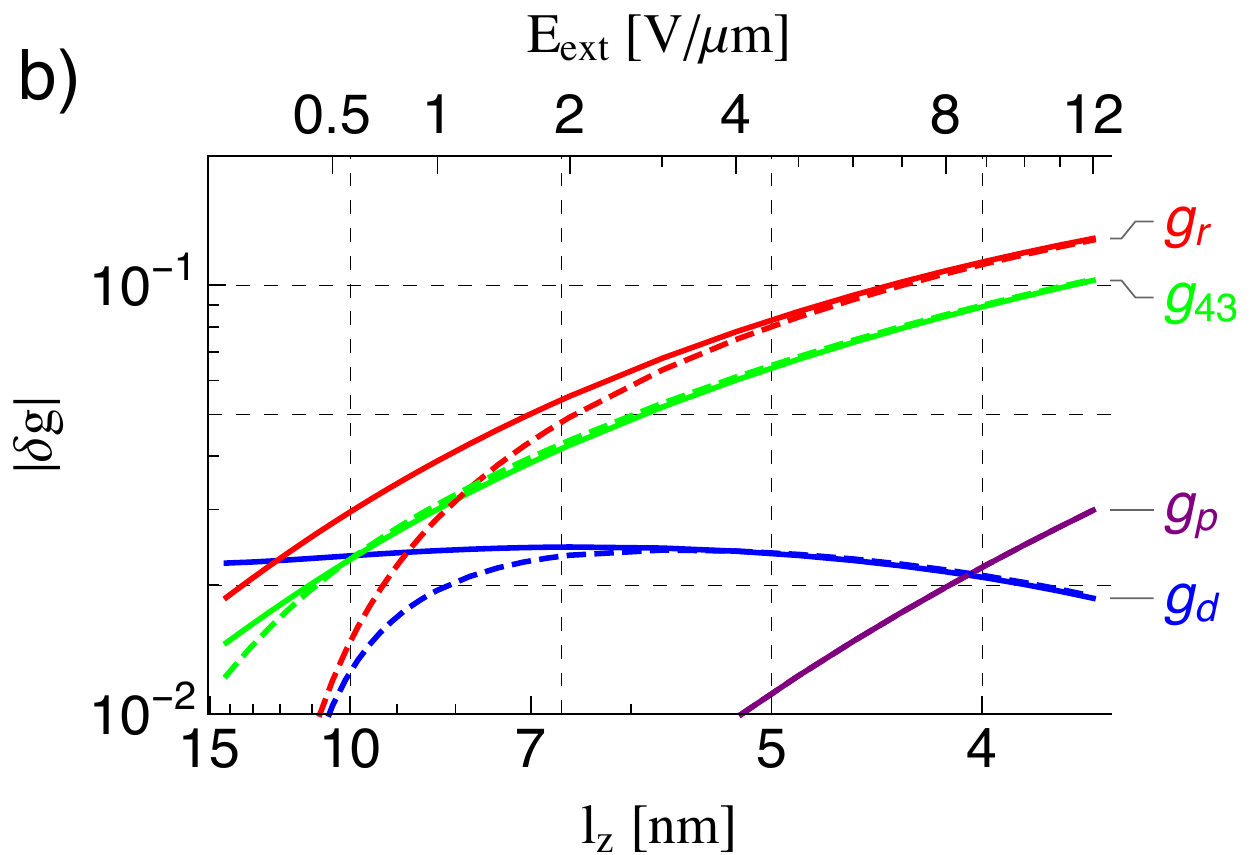}
\\\vspace{0.8cm}
\includegraphics[width=0.49\columnwidth]{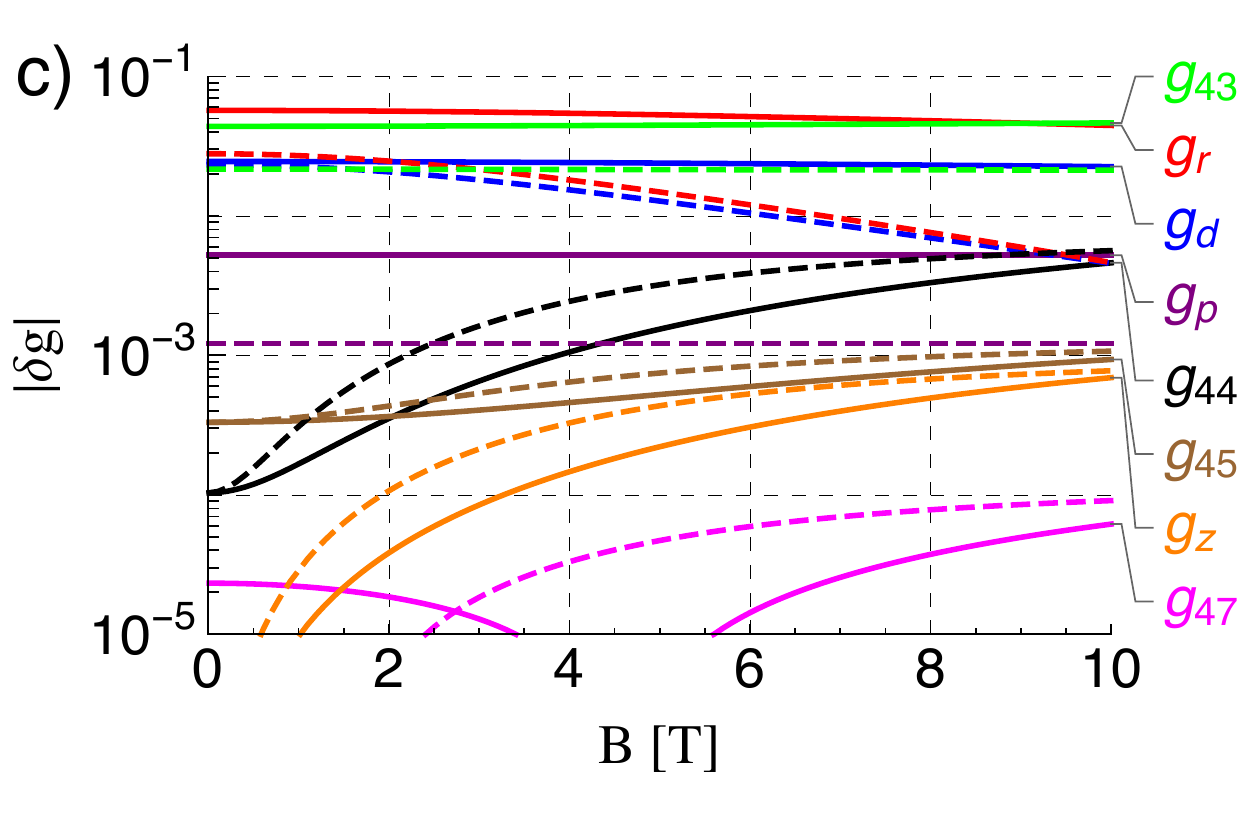}\includegraphics[width=0.49\columnwidth]{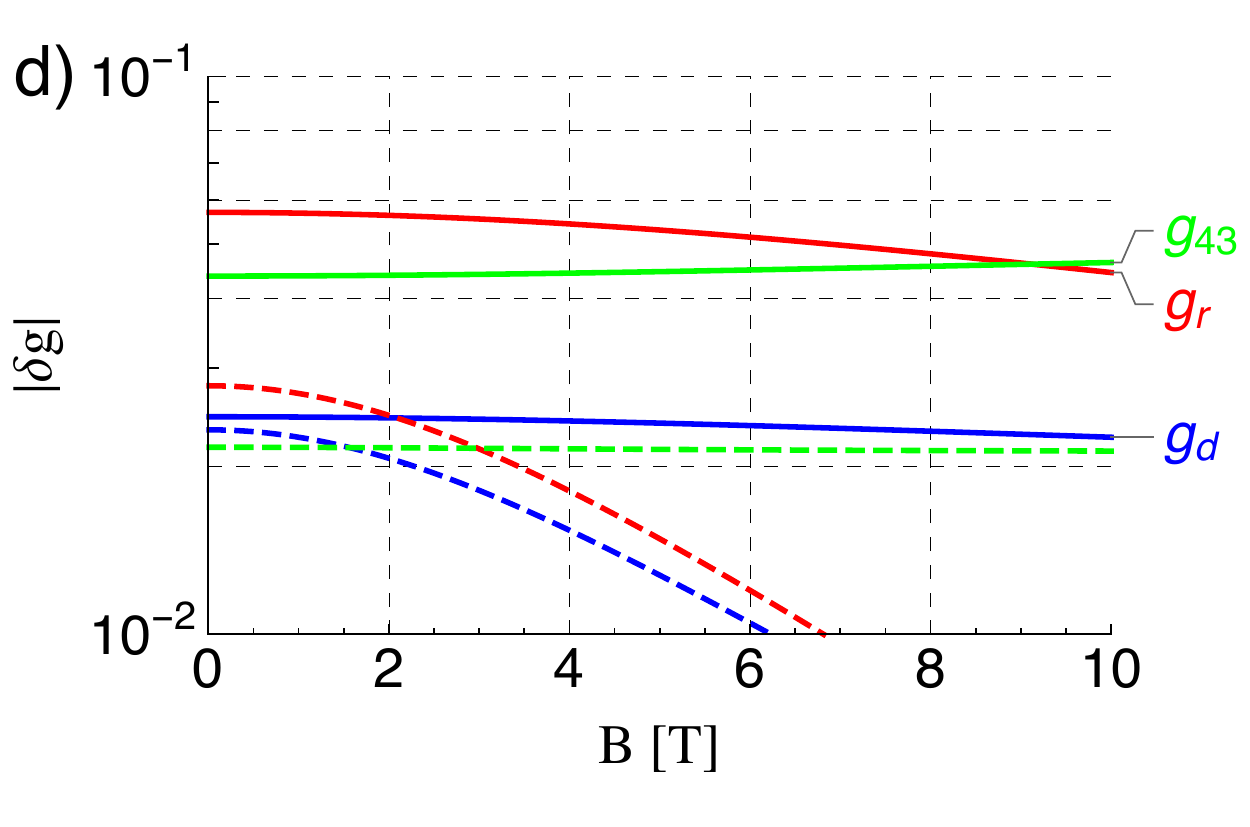}
\\\vspace{0.8cm}
\includegraphics[width=0.49\columnwidth]{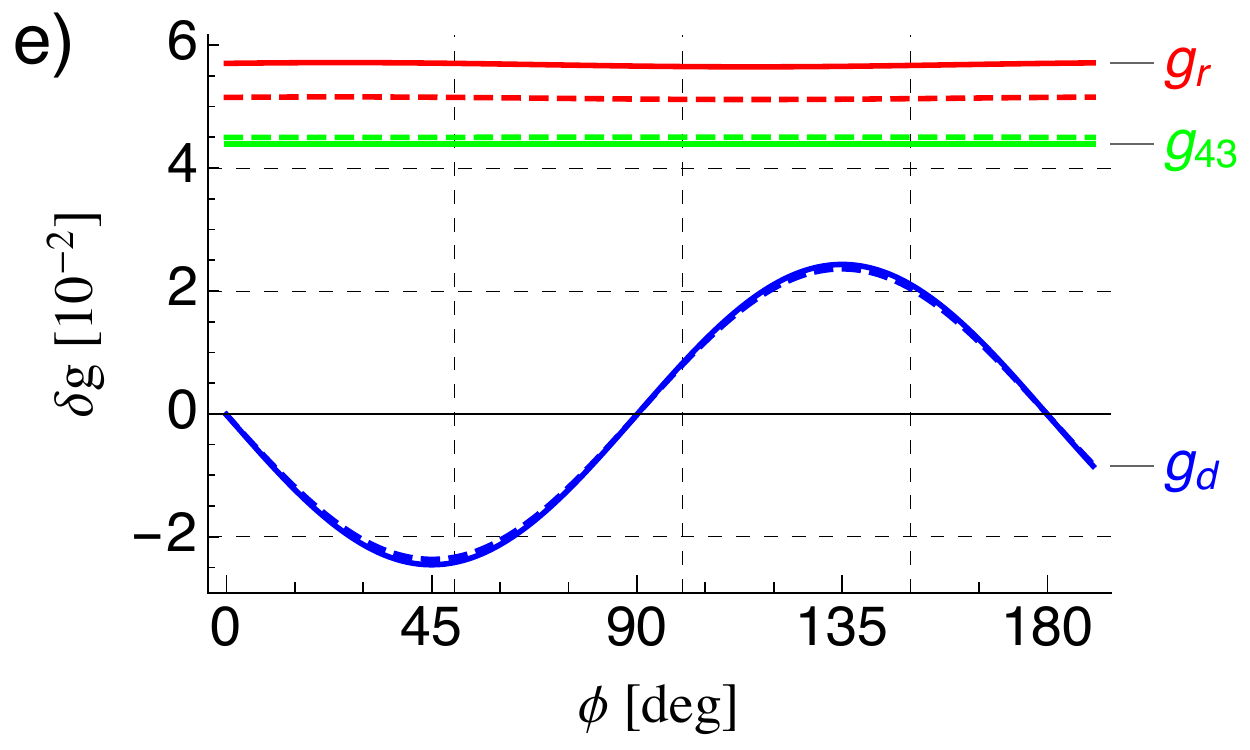}\includegraphics[width=0.49\columnwidth]{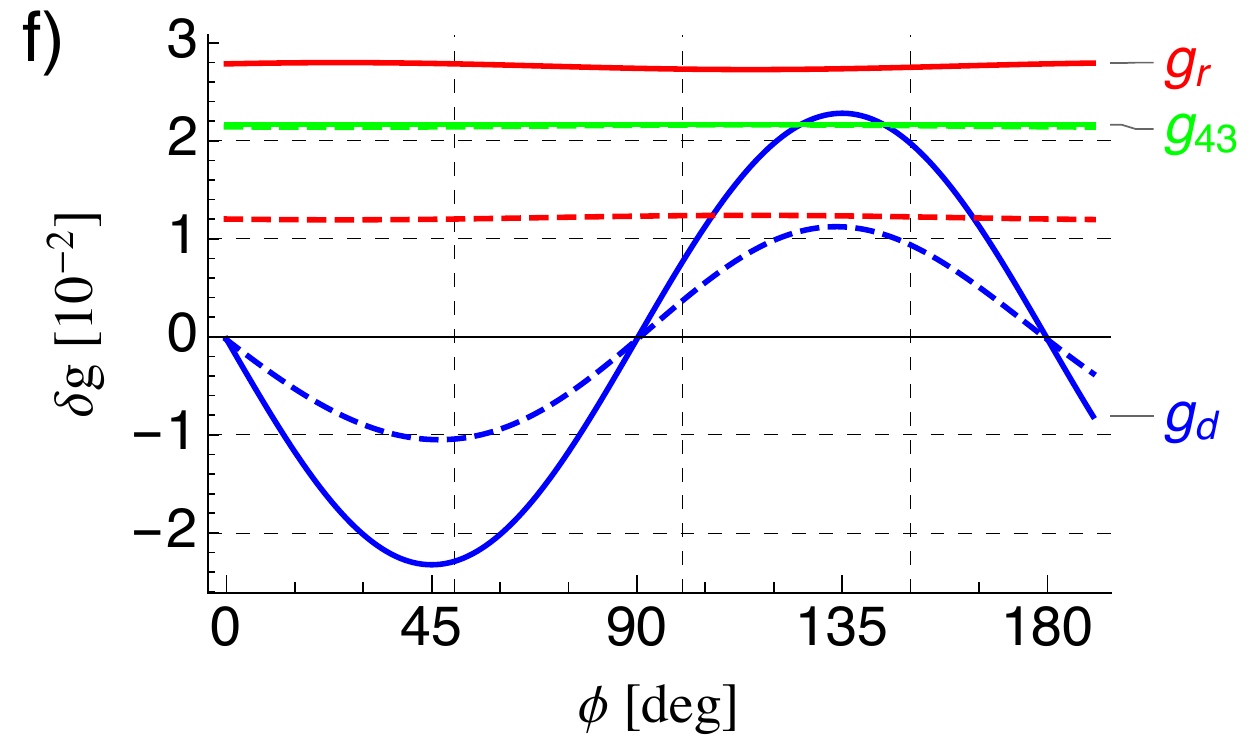}
\\\vspace{0.8cm}
\caption{\label{fig:gs}
The corrections to the g-factor labeled according to the notation of Eqs.~\eqref{eq:gd}--\eqref{eq:gz00} for the ground state $\alpha=1$, $n_x=0$, and $n_y=0$. The parameters used in this figure are $E_x=2.34$ meV, $E_y=2.61$ meV, $\delta=25^\circ$, $\phi=45^\circ$, unless stated otherwise (the adopted parameters were taken from fits to data measured in Ref.~\citenum{camenzind1}, see Fig.~7 in Ref.~\citenum{stano2017}). (a-b) As a function of the 2DEG width, parametrized by the nominal width $l_z$ (the lower x axis) and the interface electric field $E_{\rm ext}$  (the upper x axis). The solid (dashed) curves show corrections for $B = 0$~T ($B=6$ T).  
Panel (b) shows the same as (a) apart from the y-axis range.
(c-d) As a function of the magnetic field. The solid (dashed) curves show corrections for $E_{\rm ext}=2.14$ V/$\mu$m ($E_{\rm ext}=0.5$ V/$\mu$m). 
Panel (d) shows the same as (c) apart from the y-axis range.
(e-f) As a function of the magnetic field orientation for (e) $E_{\rm ext}=2.14$ V/$\mu$m (corresponding to $\lambda_z=6.5$ nm) and (f) $E_{\rm ext}=0.5$ V/$\mu$m (corresponding to $\lambda_z=10.5$ nm). The solid (dashed) curves show corrections for $B=0$ T ($B=6$ T).  In plotting these figures, we used formulas in Eqs.~\eqref{eq:gd}--\eqref{eq:gz00}. In calculating expression $g_{\textrm{x},2}$, for $\mathrm{x}=\mathrm{d,r,\ldots}$, we use the replacement $\Phi^2 \to 1-1/(1+\Phi^2)$, see Eq.~(41) in Ref.~\citenum{stano2017}, to regularize the unphysical divergence for $\Phi \geq 1$.
}
\end{figure}

\begin{figure}
\includegraphics[width=0.49\columnwidth]{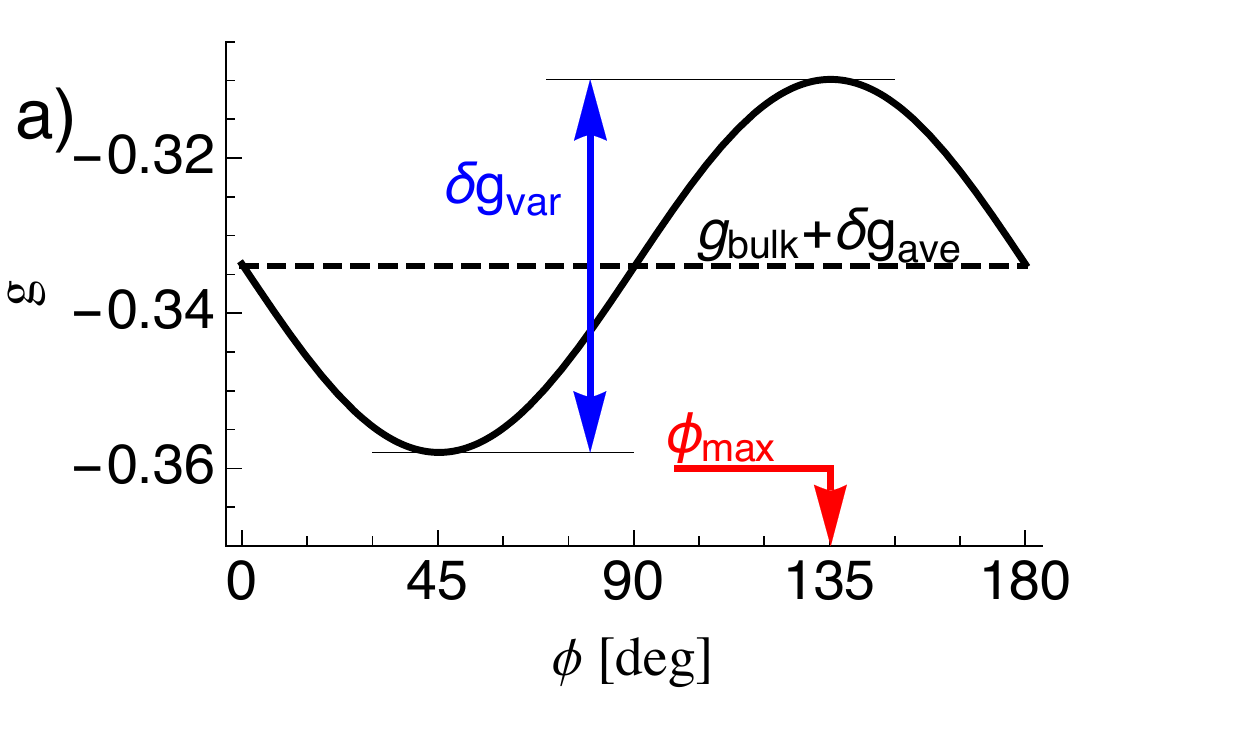}
\includegraphics[width=0.49\columnwidth]{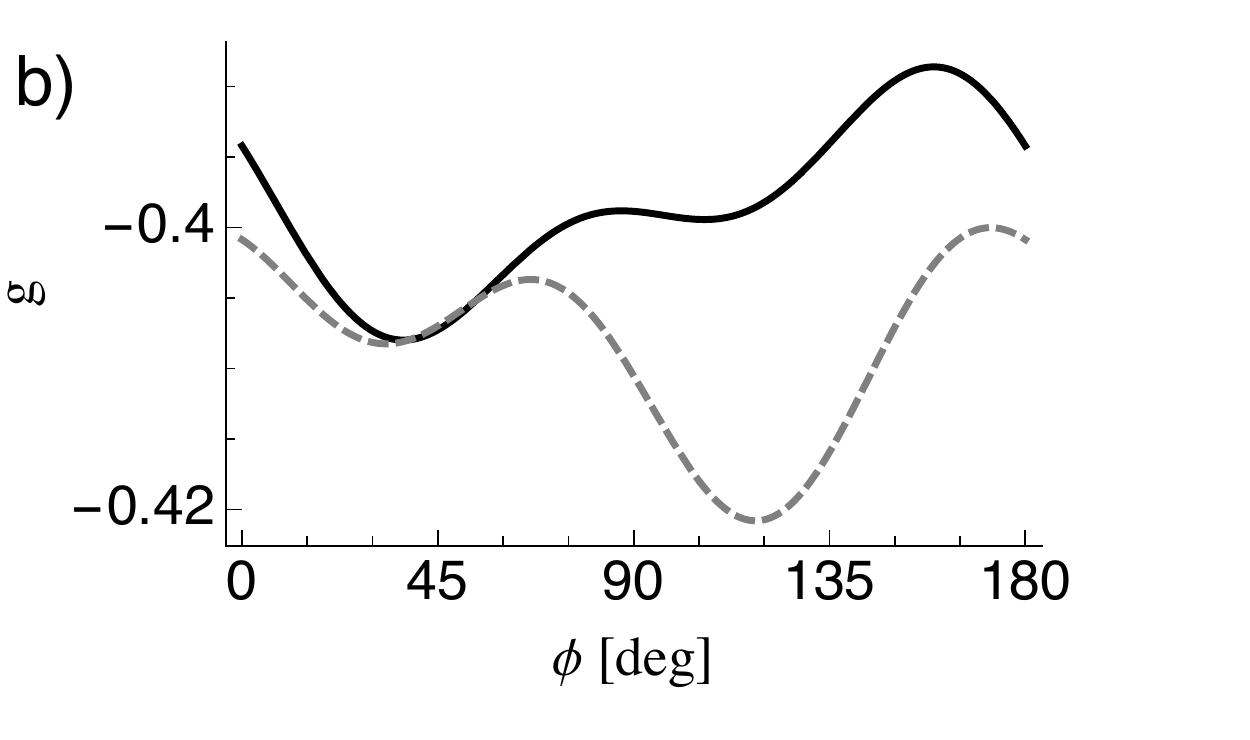}
\\\vspace{0.5cm}
\includegraphics[width=0.49\columnwidth]{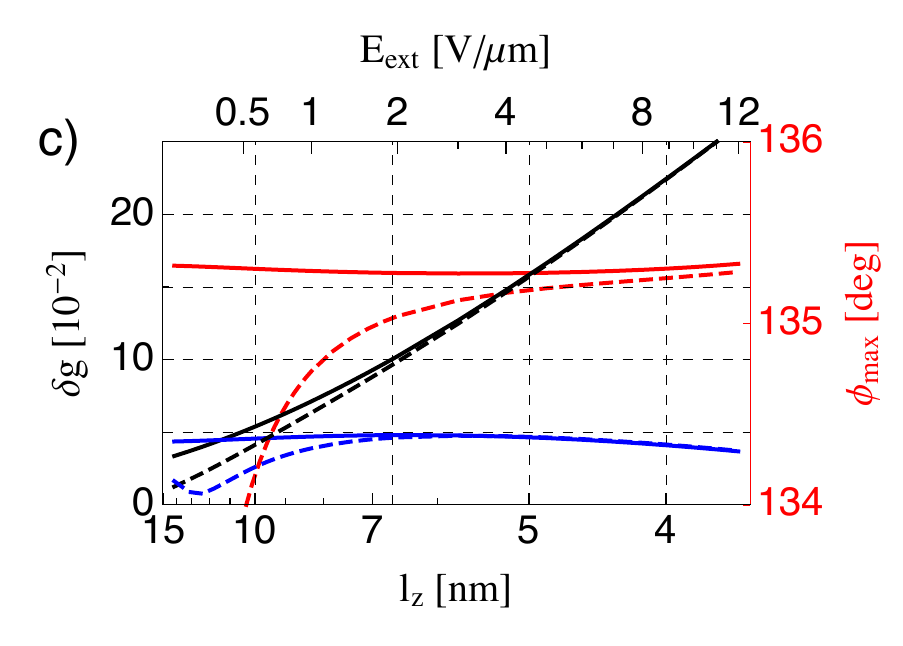}
\includegraphics[width=0.49\columnwidth]{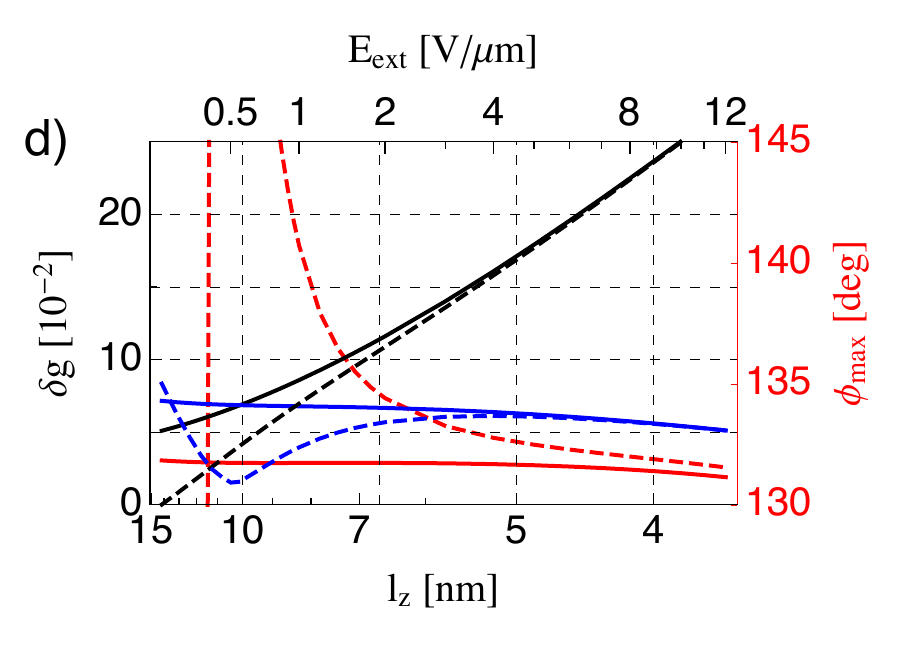}
\\\vspace{0.5cm}
\includegraphics[width=0.49\columnwidth]{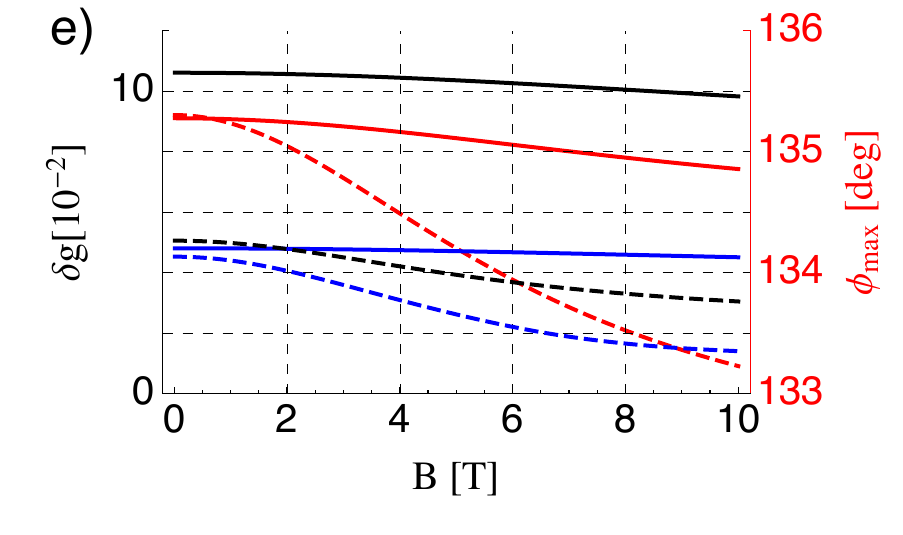}
\includegraphics[width=0.49\columnwidth]{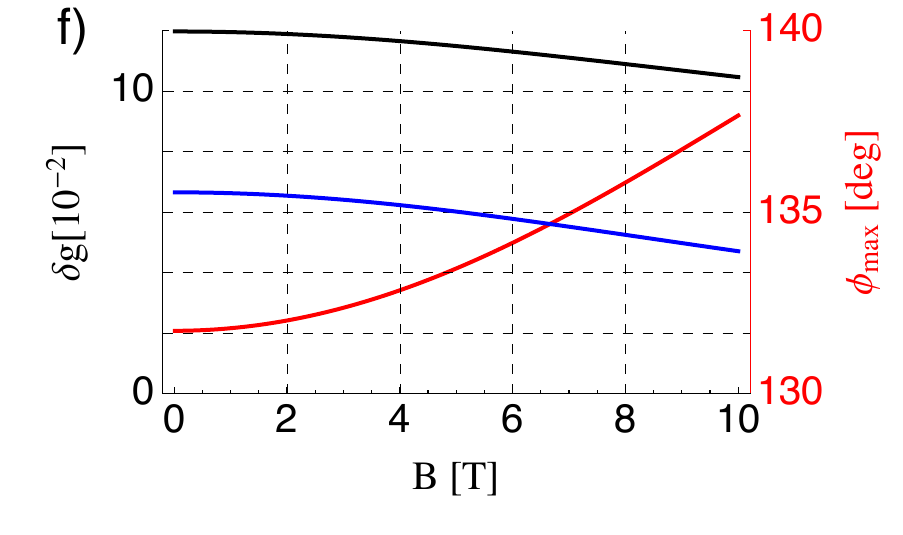}
\\\vspace{0.5cm}
\caption{\label{fig:gtot}
The total g-factor (or the g-factor correction) for the ground state $\alpha=1$, $n_x=0$, and $n_y=0$ [panels (a), (c), and (e)] and the excited state $\alpha=1$, $n_x=1$, and $n_y=0$ [panels (b), (d), and (f)] for $E_x=2.34$ meV, $E_y=2.61$ meV, $\delta=25^\circ$, and $E_{\rm ext}=2.14$ V/$\mu$m (unless stated otherwise). (a) The ground state g-factor as a function of the magnetic field direction. (b) The excited state g-factor for $B=5$ T (solid) and $B=7$ T (dashed). Panel (a) represents the most typical case: the g-factor is a curve that is well described by its average over $\phi$ [black; we also subtract the constant -0.44 from it when plotting it in panels (c)-(f)], variation (blue; defined as the difference of the maximal and minimal value as a function of $\phi$) and the magnetic-field orientation at the maximum (red). These three quantities are plotted in panels (c)-(f) in the corresponding colors. (c) As a function of the 2DEG width for $B = 0$ (solid) and $B=6$ T (dashed). (d) Same as (c) for the excited state.
(e) As a function of the magnetic field for $E_{\rm ext}=2.14$ V/$\mu$m (solid) and $E_{\rm ext}=0.5$ V/$\mu$m (dashed). (f) Same as (e) for the excited state. In panel (f) we do not plot quantities for the smaller interface field. The reason is that for small interface field (wide 2DEG) and high magnetic field, the g-factor for excited states does not typically look like the curve plotted in panel (a), but rather like in panel (b). More quantities would be needed to characterize the function, which we do not do for clarity of the figures. One should resort to full formulas in this case.
}
\end{figure}

\end{widetext}

\section{Conclusions}

In this article, we have analyzed in-plane-magnetic-field induced terms in the Hamiltonian describing an electron confined in a quasi-two-dimensional quantum dot. We have focused on terms that can be grasped by restricting this Hamiltonian effectively only to the spin degree of freedom,
\be
H^{(\alpha,i)}_{\rm eff} = \frac{\mu_B}{2} \mathbf{B} \cdot \mathbf{g}_{\alpha,i} \cdot\boldsymbol{\sigma},
\label{eq:summary}
\ee 
meaning that the orbital degrees of freedom of the electron are fixed to be the subband $\alpha$ and the in-plane orbital $i$. The g-tensor grasps all the spin-related properties of the electron under such approximation. Apart from the explicit dependence on the quantum numbers $\alpha$ and $i$, the g-tensor depends on the heterostructure confinement shape and strength, and the magnetic field magnitude and orientation. We have analyzed these dependences in great details.\footnote{We have assumed a spatially constant magnetic field in this article. In quantum-dot devices, a micromagnet is often incorporated, inducing an additional, spatially non-uniform, magnetic field. In the lowest-order approximation, this situation would be grasped by replacing $\mathbf{B}$ by $\mathbf{B}({\bf r_0})$ in Eq.~\eqref{eq:summary}, where $\mathbf{r}_0$ is the position of the quantum dot and is a function of the applied electric fields. In the next order, there will be effects arising from cross-terms including the magnetic field gradients and the spin-orbit interactions. We did not cover such effects here.}

The various ``spin-orbit'' interactions of the bulk zinc-blende crystal are the microscopic origin of the g-tensor corrections that we calculated here. Importantly, these interactions go beyond the most usually considered Rashba and Dresselhaus terms. For example, the time-reversal-antisymmetric term denoted as $H_{43}$, see Eq.~\eqref{eq:H43}, dominates the g-tensor corrections for typical parameters of GaAs/AlGaAs quantum dots. The ``standard'' Rashba spin-orbit term also gives a sizable contribution and the Dresselhaus term dominates the directional anisotropy (the variation of the g-tensor upon changing the magnetic field direction in the 2DEG plane).

We make specific predictions that can be tested experimentally, for example as for the directional anisotropies or magnetic field dependence of the Zeeman energy of the quantum dot with a single electron. Taking an alternative view, these predictions allow to extract several $k\cdot p$ constants from such measurements. Finally, our results have direct implications for electrical manipulation schemes of spin qubits and for understanding of their susceptibility to electrical noise.

\acknowledgments

This work was supported by JSPS Kakenhi Grant No. 16K05411, CREST JST (JPMJCR1675), the Swiss National Science Foundation, and the NCCR QSIT.

\appendix

\section{List of the off-diagonal effective Zeeman terms \label{app:off-diagonal}}

Here, we list the off-diagonal g-tensor terms magnitude, up to the third order in the in-plane field. The terms arising from the Dresselhaus spin-orbit interaction are given by
\begin{subequations}
\label{eq:gd_perp}
\begin{eqnarray}
g_{\rm d,0}^{\perp}&=& \frac{\lambda_d}{\lambda_z} \Big( c_1 \cos (2\phi) + c_2 \eta_+ \cos (2\phi) -c_2 \eta_- \cos (2\delta) \Big),\nonumber \\ \label{eq:gd0_perp}\\
g_{\rm d,2}^{\perp}&=& \frac{\lambda_d}{\lambda_z} \Phi^2  \Big( c_5 \cos (2\phi) + (c_6-c_{14}) \eta_+ \cos (2\phi) \nonumber \\ 
&&\qquad \qquad + c_{14} \eta_- \cos (2\delta)\nonumber\\
&&\qquad \qquad -c_4 \eta_- \sin (2\phi) \sin(2\phi-2\delta) \label{eq:gd2_perp} \\ 
&&\qquad \qquad -c_6 \eta_- \cos (2\phi) \cos(2\phi-2\delta) \nonumber \\ 
&&\qquad \qquad - c_1 c_{16} \cos(2\phi) [\eta_+ - \eta_- \cos (2\phi-2\delta)] \nonumber
\Big), 
\end{eqnarray}
\end{subequations}
whereas $g_{\rm d,1}^{\perp}=0$. 
The Rashba spin-orbit interaction gives rise to 
\begin{subequations}
\label{eq:gr_perp}
\begin{eqnarray}
g_{\rm r,0}^{\perp} &=& - 4 \xi_{r} c_{11} \eta_{-} \sin(2\phi-2\delta) ,  \label{eq:gr0_perp} \\
g_{\rm r,2}^{\perp} &=&  \xi_{r} \Phi^2 c_{15} \eta_{-} \sin(2\phi-2\delta),  \label{eq:gr2_perp}
\end{eqnarray}
\end{subequations}
with $g_{\rm r,1}^{\perp} = 0$.
The $H_{43}$ terms do not give the off-diagonal g-tensor components, $g_{43,0}^{\perp} = g_{43,1}^{\perp} = g_{43,2}^{\perp} =0$. 
The $H_{44}$ terms lead to 
\begin{subequations}
\label{eq:g44_perp}
\begin{eqnarray}
g_{44,0}^{\perp} &=& - \frac{\lambda_{44}^2}{\lambda_z^2} \eta_- \cos(2\phi) \sin (2\delta) ,  \label{eq:g440_perp} \\
g_{44,2}^{\perp} &=& \frac{\lambda_{44}^2}{\lambda_z^2} \Phi^2 \Big( \frac{c_{18}}{4} \sin (4\phi)  + 2c_{19} \cos(2\phi) \nonumber \phantom{xxxxxx}\\
&& \hspace{0.6in} \times  [\eta_+ \sin(2\phi) - \eta_- \sin (2\delta)] \Big),  \label{eq:g442_perp}
\end{eqnarray}
\end{subequations}
with $g_{44,1}^{\perp} = 0$. The terms from $H_{45}$ give 
\begin{subequations}
\label{eq:g45_perp}
\begin{eqnarray}
g_{45,0}^{\perp} &=& \frac{\lambda_{45}^2}{\lambda_z^2} \eta_- \sin(2\phi) \cos (2\delta)  ,  \label{eq:g450_perp} \\
g_{45,2}^{\perp} &=& -\frac{\lambda_{45}^2}{\lambda_z^2} \Phi^2 \Big( \frac{c_{18}}{4} \sin (4\phi) + 2c_{19} \sin(2\phi) \nonumber \phantom{xxxxxx}\\
&& \hspace{0.6in} \times  [\eta_+ \cos(2\phi) - \eta_- \cos (2\delta)] \Big),  \label{eq:g452_perp}
\end{eqnarray}
\end{subequations}
with $g_{45,1}^{\perp} = 0$.
The $H_{47}$ term gives rise to,
\begin{subequations}
\label{eq:g47_perp}
\begin{eqnarray}
g_{47,0}^{\perp} &=& \Big( -\frac{\lambda_{47}^3}{l_{z}^3} + \frac{\lambda_{47}^{\prime}}{\lambda_{z}} c_{22} \Big) \cos(2\phi),   \label{eq:g470_perp} \\
g_{47,2}^{\perp} &=& \frac{\lambda_{47}^{\prime}}{\lambda_{z}}  \Phi^2  \cos(2\phi)  \Big(  c_{23} + (3 c_{24} -c_{22} c_{16}) \phantom{xxxxxx}\nonumber \\
&& \hspace{0.7in} \times   [\eta_+  - \eta_- \cos (2\phi-2\delta)] \Big), 
  \label{eq:g472_perp}
\end{eqnarray}
\end{subequations}
with $g_{47,1}^{\perp} = 0$.
The interface terms do not lead to off-diagonal terms, 
$g_{\rm z,0}^{\perp} = g_{\rm z,1}^{\perp} = g_{\rm z,2}^{\perp} =0$.

\section{Derivation of the spin-dependent corrections~\label{app:spin-dep}}

Here, we derive Eqs.~\eqref{eq:gd}--\eqref{eq:gi} and Eqs.~\eqref{eq:gd_perp}--\eqref{eq:g47_perp}. To this end, we aim at computing the corrections to the spin Hamiltonian for a chosen orbital state $|\alpha i\rangle$, applying up to the third order perturbation theory (see Footnote 1 in Ref.~\citenum{stano2017} for a comment on the name of this method). The perturbation comprises the spin and the magnetic-field originated terms,
\be
H^\prime = H^\prime_S + H^\prime_B.
\ee
In the following, we first derive the third-order perturbative formula, adjusting the general theory for our case. Then, we derive the Zeeman-term corrections, taking the constituents of $H^\prime_S$ one by one. Before that, we note that, as explained in the discussion around Eq.~\eqref{eq:split}, we will express these corrections as a sum of two terms,
\begin{subequations}
\be
\delta V_{||} \sigma_{||} + \delta V_{\perp} \sigma_{\perp}.
\label{eq:v-vector}
\ee
It means that we separate the components parallel and perpendicular to the in-plane magnetic field,
\begin{eqnarray}
\sigma_{||} &\equiv& \boldsymbol{\sigma} \cdot {\bf B}/|{\bf B}| = \sigma_x\cos \phi  + \sigma_y \sin \phi, \\
\sigma_{\perp} &\equiv& \boldsymbol{\sigma} \cdot \left( {\bf B}/|{\bf B}| \times \z \right) = \sigma_x\sin \phi - \sigma_y \cos \phi.
\end{eqnarray}
\end{subequations}
The parallel components $\delta V_{||}$, leading to Eqs.~\eqref{eq:gd}--\eqref{eq:gi}, change the Zeeman energy. The perpendicular components $\delta V_{\perp}$, leading to Eqs.~\eqref{eq:gd_perp}--\eqref{eq:g47_perp}, change (slightly rotate) the eigenspinor direction.

\subsection{Perturbation theory up to the third order}
\label{app:Lowdin}

In the subspace defined by subband $\alpha$ and orbital state $i$, the effective Hamiltonian up to the third order is
\be
\begin{split}
& \langle \alpha i | H^\prime | \alpha i\rangle +
{\sum_{\beta j \neq \alpha i}} \frac{\langle \alpha i | H^\prime | \beta j\rangle \langle \beta j | H^\prime | \alpha i\rangle } {E_{\alpha i}-E_{\beta j}} \\
& + \sum_{\beta j \neq \alpha i} \sum_{\gamma k \neq \alpha i}  \frac{ \langle \alpha i | H^\prime | \beta j \rangle\langle \beta j | H^\prime | \gamma k \rangle \langle \gamma k | H^\prime | \alpha i \rangle  }{(E_{\alpha i}-E_{\beta j})(E_{\alpha i}-E_{\gamma k})}   \\
&  - \frac{1}{2}  \sum_{\beta j \neq \alpha i} \frac{\{ \langle \alpha i | H^\prime | \beta j \rangle\langle \beta j | H^\prime | \alpha i \rangle, \langle \alpha i | H^\prime | \alpha i \rangle \}}{(E_{\alpha i}-E_{\beta j})^2}.
\end{split}
\label{eq:Lowdin-3rd}
\ee
The first two terms correspond to Eq.~\eqref{eq:Lowdin} for $i=j$, the next two terms arise in the third order. In deriving this expression from the general formulas (see, for example, page 135 in Ref.~\citenum{birpikus}), we used that since the Zeeman term is included among the perturbations in $H^\prime$, the basis state energies are spin independent. It allows us to suppress the spin indexes, by treating the matrix elements such as $\langle \alpha i | H^\prime | \beta j\rangle$ as operators in the spin space (that is, two-by-two matrices). This is why the anticommutator in the last term is necessary. 

We now simplify the third order terms further, by restricting to contributions which are linear in the spin-dependent perturbation $H^\prime_{\rm S}$ and up to the third order in the magnetic field. 
From $\langle \alpha i | H^\prime_1 | \alpha i\rangle=0$ and $\langle \alpha i | H^\prime_2 | \beta j\rangle=0$ if $j\neq i$, it follows that the last line in Eq.~\eqref{eq:Lowdin-3rd} can be written as
\be
\begin{split}
& - \sum_{\beta \neq \alpha} \frac{\langle i | \ave{H^\prime_1}{\alpha\beta} \ave{H^\prime_1}{\beta\alpha} | i \rangle}{(E_{\alpha}-E_{\beta})^2}
 \langle \alpha i | H_{\rm S}^\prime | \alpha i \rangle \\
& - \sum_{\beta \neq \alpha} \frac{\langle i | ( \ave{H^\prime_1}{\alpha\beta} \ave{H^\prime_{\rm S}}{\beta\alpha} + \ave{H^\prime_{\rm S}}{\alpha\beta} \ave{H^\prime_1}{\beta\alpha} ) | i \rangle }{ (E_{\alpha}-E_{\beta})^2 }
 \langle \alpha i | H_2^\prime | \alpha i \rangle.
\end{split}
\ee
The anticommutator is not needed anymore since $H^\prime_1$ and $H^\prime_2$ do not contain Pauli matrices.
The first line of the above formula can be considered as a correction to the intra-subband contribution, such as the one in Eq.~(\ref{eq:example1}). The second line results in corrections with the same angular dependence as the inter-subband contributions, such as the one in Eq.~(\ref{eq:example2}).
As a result, we split the effective Hamiltonian to the following two contributions,
\begin{subequations}
\label{eq:Lowdin_spin}
\be 
\begin{split}
\delta H_{\rm s}^{(1)} =& \langle i | \ave{H_{\rm S}^\prime}{\alpha}  | i\rangle (1-\xi_{\alpha i}) +
\langle i | [\ave{H_{\rm S}^\prime}{\alpha} , \frac{e}{i \hbar} \ave{ \Ain  }{\alpha} \cdot {\bf r}] | i\rangle \\
& + \frac{1}{2} \langle i | [ [\ave{H_{\rm S}^\prime}{\alpha} , \frac{e}{i \hbar} \ave{ \Ain  }{\alpha} \cdot {\bf r}], \frac{e}{i \hbar} \ave{ \Ain  }{\alpha} \cdot {\bf r}]  | i\rangle,
\end{split}
\label{eq:spin_intra2} 
\ee
obtained with the help of the identity $H^\prime_1= (e/ i\hbar) \left[ \Ain \cdot \Rin ,h_{2D}  \right]$, and
\be 
\begin{split}
\delta  H_{\rm s}^{(2)} =& {\sum_{\beta\neq \alpha}} \sum_{k}   \frac{\langle i | \ave{H^\prime}{\alpha\beta}  | k\rangle \langle  k | \ave{H^\prime}{\beta\alpha}  | i\rangle}{E_{\alpha}-E_{\beta}} \\
& \hspace{-1.0cm} +\sum_{\beta \neq \alpha}\sum_{\gamma \neq \alpha} \sum_{j, k}  \frac{ \langle i | \ave{H^\prime}{\alpha\beta}  | j \rangle\langle j | \ave{H^\prime}{\beta\gamma}  | k \rangle \langle k | \ave{H^\prime}{\gamma\alpha}  | i \rangle  }{(E_{\alpha}-E_{\beta})(E_{\alpha}-E_{\gamma})} \\
& \hspace{-1.0cm} - \ave{ H_2^\prime }{\alpha}  \sum_{\beta \neq \alpha} \frac{\langle i | \ave{H^\prime_1}{\alpha\beta} \ave{H^\prime_{\rm S}}{\beta\alpha} + \ave{H^\prime_{\rm S}}{\alpha\beta} \ave{H^\prime_1}{\beta\alpha}| i \rangle}{(E_{\alpha}-E_{\beta})^2} .
\label{eq:spin_inter}
\end{split}
\ee
\end{subequations}
The correction factor in the first is
\be
\xi_{\alpha i} = \sum_{\beta\neq \alpha} \frac{\langle i | \ave{H^\prime_1}{\alpha\beta} \ave{H^\prime_1}{\beta\alpha} | i \rangle}{(E_{\alpha}-E_{\beta})^2}
\label{eq:xi}
\ee
and we neglected the orbital with respect to the subband excitation energies in the second.

Obviously, the role of the commutators is to assure gauge invariance. After demonstrating it in Section \ref{subsec:H_D intraband}, we put $z_0=\ave{z}{\alpha}$, upon which the commutators disappear since $\ave{\Ain}{\alpha}=0$. We denote the gauge invariant form of the expectation value in Eq.~\eqref{eq:spin_intra2}, without the correction $\xi_{\alpha i}$, as the intra-subband contribution. The terms from Eq.~\eqref{eq:spin_inter} are denoted as the inter-subband terms. The correction term $\xi_{\alpha i}$ is calculated separately, in Appendix \ref{app:cs}. Finally, we note that $\delta H_{\rm s}$ is still an operator in the spin space, and it can be written in the form of Eq.~(\ref{eq:v-vector}). 
In the following subsections, we compute the intra-subband and inter-subband contributions for every constituent of $H^\prime_S$ separately. From here on, we assume zero out-of-plane component of the magnetic field, $B_z=0$. 

Before continuing, we note that the fourth-order perturbation theory will generate additional corrections that are cubic in the in-plane magnetic flux. These corrections contain one matrix element from the zero-field Rashba or Dresselhaus (i.e. $H_{\rm d, 0}$) spin-orbit interactions, and three matrix elements from $H_1^{\prime}$. However, since each of these matrix elements comprises an in-plane momentum operator $p_x$ or $p_y$, the corrections will be of order $O(\Phi^3) \times O(\eta_{\pm}^2)$. Since this is of the same or higher order of magnitude as the errors introduced when we neglect the in-plane orbital splitting in deriving Eq.~\eqref{eq:spin_inter}, in what follows we also neglect the fourth-order terms.

\subsection{Dresselhaus intra-subband terms \label{subsec:H_D intraband}}

We first take Eq.~\eqref{eq:spin_intra2} with the Dresselhaus term as the perturbation, $H_{\rm S}^\prime \rightarrow H_{\rm D}$, and sort the resulting terms according to the power of the magnetic field.
The term independent on the in-plane magnetic field is the unperturbed Dresselhaus interaction, Eq.~\eqref{eq:Dresselhaus02}. The linear term is
\be \begin{split}
\delta H_{\rm d,1}^{(1)} =& \frac{e\gamma_c}{2\hbar^3}  \left( \ave{\{ \sigma_{y}\Ainy  - \sigma_{x} \Ainx , p_z^2 \} }{\alpha} 
\right. \\
&\left. - \{ \sigma_{y}\ave{\Ainy }{\alpha} - \sigma_{x} \ave{\Ainx }{\alpha}, \ave{p_z^2}{\alpha} \} \right),
\end{split} \ee
where we have used that the expectation values of the momentum operators, $p_x$, and $p_y$, are zero in any localized state, including state $i$. Using Eq.~(\ref{eq:Ain}) further gives
\be \begin{split}
\delta H_{\rm d,1}^{(1)} = &-\frac{e\gamma_c}{2\hbar^3} \left( B_{x} \sigma_{y} + B_{y} \sigma_{x} \right) \\
& \times \left( \ave{\{ z-z_0, p_z^2 \} }{\alpha} - \{ \ave{z-z_0}{\alpha} , \ave{p_z^2}{\alpha} \}\right). 
\end{split} \ee
The gauge choice $z_0 =\ave{z}{\alpha}$ simplifies it further,
\be
\delta H_{\rm d,1}^{(1)} = -\frac{e\gamma_c}{2\hbar^3} \left( B_{x} \sigma_{y} + B_{y} \sigma_{x} \right) \ave{\{ \Delta z, p_z^2 \} }{\alpha},
\ee
with $\Delta z = z - \ave{z}{\alpha}$.
The components of ${\bf \delta V}$ follow as
\begin{subequations}
\begin{eqnarray}
\delta V_{\rm d,1,||}^{(1)}  &=& -\frac{e\gamma_c}{2 \hbar^3} B \sin (2\phi) \ave{\{ \Delta z, p_z^2 \} }{\alpha} , \\
\delta V_{\rm d,1,\perp}^{(1)}  &=& \frac{e\gamma_c}{2 \hbar^3}  B \cos (2\phi)  \ave{\{ \Delta z, p_z^2 \} }{\alpha},
\end{eqnarray}
\end{subequations}
which correspond to the terms proportional to $c_1$ in Eqs.~\eqref{eq:gd0}, and \eqref{eq:gd0_perp}, respectively.

The term quadratic in the in-plane field is
\be \begin{split}
\delta H_{\rm d,2}^{(1)} = & \frac{e^2\gamma_c}{2\hbar^3} \sigma_{z} \left( B_{y}^2 - B_{x}^2 \right) 
\left( \ave{\{ \Delta z^2, p_z \} }{\alpha} + \right. \\
&\left. +\{ ( \ave{ \Delta z}{\alpha})^2 , \ave{p_z}{\alpha} \} - 2\ave{\{ \Delta z, p_z \} }{\alpha} \ave{ \Delta z }{\alpha}
\right),
\end{split} \ee
which is again simplified taking $z_0 =\ave{z}{\alpha}$ to
\be
\delta H_{\rm d,2}^{(1)} = \frac{e^2\gamma_c}{2\hbar^3} \sigma_{z} \left( B_{y}^2 - B_{x}^2 \right) 
  \ave{\{ \Delta z^2, p_z \} }{\alpha}.
\ee
Since the last term is an expectation value of a purely imaginary operator, this correction is zero. 
 
Following the same procedure, the cubic term gives 
\begin{subequations}
\begin{eqnarray}
\delta V_{\rm d,3,||}^{(1)}  &=&\frac{e^3\gamma_c}{2 \hbar^3} B^3 \sin(2\phi) \ave{ \Delta z^3 }{\alpha} ,
\\ \delta V_{\rm d,3,\perp}^{(1)}  &=& 0,
\end{eqnarray}
\end{subequations}
corresponding to the term $c_3$ in Eq.~\eqref{eq:gd2}.

\subsection{Dresselhaus inter-subband terms \label{subsec:H_D interband}}

We now consider the Dresselhaus term in Eq.~\eqref{eq:spin_inter}, again sorting the terms according to powers of the in-plane magnetic field.
We first compute the first line of Eq.~\eqref{eq:spin_inter}, arising from the second-order perturbation theory.
We calculate the terms involving $H^\prime_1$, and $H^\prime_2$ separately, starting with the former. 
The linear term, coming from $H_{\rm d,0}$, Eq.~\eqref{eq:powers}, in the first term of Eq.~\eqref{eq:spin_inter} gives
\be
\begin{split}
\delta H_{\rm d,1}^{(2)} = & \frac{e\gamma_c}{m \hbar^3} {\sum_{\beta\neq \alpha}}  \frac{1}{E_{\alpha }-E_{\beta}}  z_{\alpha\beta} \ave{ p_z^2 }{\beta\alpha} \\
&\times \langle  i | \left[  \left( B_y p_x - B_x p_y \right) \left( -\sigma_x p_x + \sigma_y p_y \right) \right.  \\
& \hspace{0.1in} + \left. \left( -\sigma_x p_x + \sigma_y p_y \right) \left( B_y p_x - B_x p_y \right) \right]  | i \rangle ,
\end{split}
\ee
where we used the fact that $z_{\alpha\beta}$ and $\ave{ p_z^2 }{\beta\alpha}$ are real. Using the reflection symmetry of the in-plane confinement along the axes $\x_d$ and $\y_d$, we arrive at
\be
\begin{split}
\delta H_{\rm d,1}^{(2)} =&  
\frac{2e\gamma_c}{m \hbar^3} {\sum_{\beta\neq \alpha}}  \frac{1}{E_{\alpha }-E_{\beta}} z_{\alpha\beta} \ave{ p_z^2 }{\beta\alpha}  \\
&\times \left\{ \sigma_x \left[ - \ave{p_{+}^2}{i} B_y  + \ave{p_{-}^2}{i} \left( B_x \sin 2\delta -  B_y \cos 2 \delta \right) \right]\right. \\
& \hspace{0.1in} \left.  + \sigma_y  \left[ - \ave{p_{+}^2}{i} B_x  + \ave{p_{-}^2}{i} \left( B_x \cos 2\delta + B_y \sin 2 \delta \right)  \right]
\right\},
\end{split}
\ee
where we put
\be
\ave{p_{\pm}^2}{i} \equiv \frac{1}{2} \langle  i | (\Pin \cdot \x_d)^2  \pm (\Pin \cdot \y_d)^2| i \rangle.
\ee
The components of ${\bf \delta V}$ then follow as
\begin{subequations}
\begin{eqnarray}
\delta V_{\rm d,1,||}^{(2)}  &=&
\frac{2e\gamma_c}{m \hbar^3} B {\sum_{\beta\neq \alpha}}  \frac{1}{E_{\alpha }-E_{\beta}} z_{\alpha\beta} \ave{ p_z^2 }{\beta\alpha} \nonumber \\
&&\times  \left[ - \ave{p_{+}^2}{i}  \sin(2\phi)  + \ave{p_{-}^2}{i} \sin (2\delta)  \right] , \\
\delta V_{\rm d,1,\perp}^{(2)}  &=&
\frac{2e\gamma_c}{m \hbar^3} B {\sum_{\beta \neq \alpha}}  \frac{1}{E_{\alpha }-E_{\beta}} z_{\alpha\beta} \ave{ p_z^2 }{\beta\alpha} \nonumber \\
&&\times  \left[ \ave{p_{+}^2}{i}  \cos(2\phi)  - \ave{p_{-}^2}{i} \cos (2\delta)  \right],
\end{eqnarray}
\end{subequations}
which gives the $c_2$ terms in Eqs.~\eqref{eq:gd0} and \eqref{eq:gd0_perp}.

The quadratic term, coming from $H_{\rm d,1}$, is
\be
\begin{split}
\delta H_{\rm d,2}^{(2)} \approx & \frac{e^2\gamma_c}{m \hbar^3} {\sum_{\beta\neq \alpha}} \left[ \frac{z_{\alpha\beta} \ave{ \{ \Delta z,p_z\} }{\beta\alpha}}{E_{\alpha }-E_{\beta}} + \frac{ \ave{ \{ \Delta z,p_z\} }{\alpha\beta} z_{\beta\alpha}  }{E_{\alpha }-E_{\beta}}  \right] \\
&\times
\sigma_z \left[ \ave{p_{+}^2}{i} \left( B_y^2  - B_x^2 \right) + \ave{p_{-}^2}{i} \left( B_x^2  + B_y^2 \right) \cos 2 \delta  \right].
\end{split}
\ee
Since $z_{\alpha\beta}$ is real, the summand is proportional to $\ave{ \{ \Delta z,p_z\} }{\alpha\beta}+\ave{ \{ \Delta z,p_z\} }{\beta\alpha}$, which vanishes for any $(\alpha, \beta)$. As a consequence, these terms do not contribute to ${\bf \delta V}$. 

The cubic term, involving $H_{\rm d,2}$, is
\be
\begin{split}
&\delta H_{\rm d,3}^{(2)}\approx  \frac{2e^3\gamma_c}{m \hbar^3} {\sum_{\beta\neq \alpha}}  \frac{1}{E_{\alpha }-E_{\beta}} z_{\alpha\beta} \ave{ \Delta z^2 }{\beta\alpha} \\
& \times \left(\sigma_x B_x
\left\{
3 B_x B_y \ave{p_{+}^2}{i}  \right.\right. \\
& \hspace{0.5in} \left. 
- \ave{p_{-}^2}{i} \left[ (B_x^2 + 2 B_y^2 ) \sin 2\delta + B_x B_y \cos2 \delta \right]
\right\} \\
&  \hspace{0.1in}  + \sigma_y B_y \left\{ 3 B_x B_y \ave{p_{+}^2}{i} \right. \\
& \left. \left. \hspace{0.5in}  - \ave{p_{-}^2}{i}  \left[ (2B_x^2 + B_y^2) \sin 2\delta - B_x B_y \cos 2\delta \right] \right\} \right), \\
\end{split}
\ee
with $\ave{ \Delta z^2 }{\alpha\beta}$ being real. The components of ${\bf \delta V}$ follow as
\begin{subequations}
\begin{eqnarray}
\delta V_{\rm d,3,||}^{(2)}  &=&
\frac{e^3\gamma_c}{m \hbar^3} B^3 {\sum_{\beta\neq \alpha}}  \frac{1}{E_{\alpha }-E_{\beta}} z_{\alpha\beta} \ave{ \Delta z^2 }{\beta\alpha} \nonumber  \\
&&\times \left[ 3 \ave{p_{+}^2}{i}  \sin(2\phi)    -3 \ave{p_{-}^2}{i} \sin(2\delta)   \right. \nonumber \\
&& \hspace{0.1in} \left. - \ave{p_{-}^2}{i} \cos (2\phi) \sin(2\phi-2\delta)  \right],  \\
\delta V_{\rm d,3,\perp}^{(2)}  &=&
- \frac{e^3\gamma_c}{m \hbar^3} B^3 {\sum_{\beta\neq \alpha}}  \frac{1}{E_{\alpha }-E_{\beta}} z_{\alpha\beta} \ave{ \Delta z^2 }{\beta\alpha} \nonumber  \\
&&\times  \ave{p_{-}^2}{i} \sin(2\phi)  \sin(2\phi-2\delta) ,
\end{eqnarray}
\end{subequations}
giving the $c_4$ terms in Eqs.~\eqref{eq:gd2} and \eqref{eq:gd2_perp}.

We now turn to terms involving $H_2^\prime$ in the first line of  Eq.~(\ref{eq:spin_inter}). The quadratic terms in the in-plane field, arising from $H_2^\prime$ and $H_{\rm d,0}$, vanish after taking the expectation value with respect to $| i \rangle$, as they contain odd number of in-plane momentum operators. The cubic terms, due to $H_2^\prime$ and $H_{\rm d,1}$, lead to the following components of ${\bf \delta V}$,
\begin{subequations}
\begin{eqnarray}
\delta V_{\rm d,H_2,||}^{(2)}  &=& - \frac{e^3\gamma_c}{2 m \hbar^3} B^3 {\sum_{\beta\neq \alpha}}  \frac{  \ave{\Delta z^2}{\alpha\beta}  \ave{ \{ \Delta z, p_z^2 \}  }{\beta\alpha} }{E_{\alpha }-E_{\beta}}  \sin (2\phi), \nonumber \\ \\
\delta V_{\rm d,H_2,\perp}^{(2)}  &=& \frac{e^3\gamma_c}{2 m \hbar^3} B^3 
{\sum_{\beta\neq \alpha}}  \frac{  \ave{\Delta z^2}{\alpha\beta}  \ave{ \{ \Delta z, p_z^2 \}  }{\beta\alpha} }{E_{\alpha }-E_{\beta}}  
\cos (2\phi) , \nonumber \\ 
\end{eqnarray}
\end{subequations}
where we used that $\ave{ \{ \Delta z, p_z^2 \}  }{\alpha\beta}$ is real. They correspond to $c_5$ in Eqs.~\eqref{eq:gd2} and \eqref{eq:gd2_perp}.

Finally, we consider the second and third lines of Eq.~\eqref{eq:spin_inter}, arising from the third-order perturbation theory. They result in
\begin{subequations}
\begin{eqnarray}
\delta V_{\rm d, 3rd,||}^{(2)} 
&=& \frac{3\gamma_c e^3 }{m^2\hbar^3} B^3 \sum_{\beta\neq\alpha}\sum_{\gamma\neq\alpha}  \frac{z_{\alpha\beta} z_{\beta\gamma} \ave{ \Delta z p_z^2 }{\gamma\alpha}}{(E_{\alpha }-E_{\beta})(E_{\alpha }-E_{\gamma})}   \nonumber\\
&&  \times \sin (2\phi) \left[  -  \ave{p_{+}^2}{i} + \ave{p_{-}^2}{i} \cos (2\phi - 2\delta) \right] \nonumber\\ 
&& - \frac{\gamma_c e^3 }{m^2\hbar^3} B^3 \sum_{\beta\neq\alpha}  \frac{z_{\alpha\beta} \ave{ p_z^2 }{\beta\alpha} \ave{ \Delta z^2 }{\alpha}}{(E_{\alpha }-E_{\beta})^2}   \nonumber\\
&& \hspace{0.1in}  \times \left[  - \ave{p_{+}^2}{i} \sin (2\phi) + \ave{p_{-}^2}{i} \sin (2\delta) \right], \\
\delta V_{\rm d, 3rd,\perp}^{(2)} 
&=& \frac{3\gamma_c e^3 }{m^2\hbar^3}  B^3 \sum_{\beta\neq\alpha}\sum_{\gamma\neq\alpha}  \frac{z_{\alpha\beta} z_{\beta\gamma} \ave{ \Delta z p_z^2 }{\gamma\alpha}}{(E_{\alpha }-E_{\beta})(E_{\alpha }-E_{\gamma})}   \nonumber  \\
&&  \times \cos (2\phi) \left[  \ave{p_{+}^2}{i} - \ave{p_{-}^2}{i} \cos (2\phi - 2\delta) \right] \nonumber\\
&& - \frac{\gamma_c e^3 }{m^2\hbar^3} B^3 \sum_{\beta\neq\alpha}  \frac{z_{\alpha\beta} \ave{ p_z^2 }{\beta\alpha} \ave{ \Delta z^2 }{\alpha}}{(E_{\alpha }-E_{\beta})^2}   \nonumber\\
&& \hspace{0.1in}  \times \left[ \ave{p_{+}^2}{i} \cos (2\phi) - \ave{p_{-}^2}{i} \cos (2\delta) \right],
\end{eqnarray}
\end{subequations} 
giving the $c_6$ and $c_{14}$ terms in Eqs.~\eqref{eq:gd2} and \eqref{eq:gd2_perp}.

\subsection{Rashba terms \label{subsec:H_R}}

Comparing to the previous section, now the calculations are simpler as the Rashba interaction, Eq.~\eqref{eq:HR}, contains only terms of zeroth and first order in the in-plane magnetic field. The intra-subband contribution from the former is the unperturbed Rashba interaction, Eq.~\eqref{eq:Rashba0}. The latter gives
\be
\delta H_{\rm r,1}^{(1)} = - \frac{\beta_{BA} e }{\hbar} (\sigma_x B_x + \sigma_y B_y)  \ave{\delta(z)}{\alpha} \ave{ z }{\alpha},
\ee
which directly gives the $c_{10}$ term in Eq.~\eqref{eq:gr0}. Analogous terms, originating in the interface-generated  spin-orbit interactions (see Footnote \ref{fnt}), were derived in Ref.~\citenum{alekseev2013} and used to fit experiments in Si in Refs.~\citenum{ruskov2017}, \citenum{jock2018}, and \citenum{tanttu2018}.
 
We now move on to the inter-subband contributions. The term linear in the in-plane magnetic field is 
\be
\begin{split}
&\delta H_{\rm r,1}^{(2)} 
= - \frac{4e \beta_{BA} }{m\hbar}
 {\sum_{\beta\neq \alpha}}  \frac{z_{\alpha\beta} \ave{\delta(z)}{\beta\alpha}}{E_{\alpha }-E_{\beta}} \\
 &\times \left\{ \sigma_x \left[ - \ave{p_{+}^2}{i} B_x  + \ave{p_{-}^2}{i} \left( B_x \cos 2\delta + B_y \sin 2 \delta \right) \right]\right. \\
& \hspace{0.1in} \left.  + \sigma_y  \left[ - \ave{p_{+}^2}{i} B_y  - \ave{p_{-}^2}{i} \left( B_y \cos 2\delta - B_x \sin 2 \delta \right)  \right]
\right\},
\end{split}
\ee
and gives the following components of ${\bf \delta V}$,
\be
\begin{split}
\delta V_{\rm r,1,||}^{(2)}
=& - \frac{4e \beta_{BA} }{m\hbar} B 
 {\sum_{\beta\neq \alpha}}  \frac{z_{\alpha\beta} \ave{\delta(z)}{\beta\alpha} }{E_{\alpha }-E_{\beta}}  \\
 &\times \left[  -\ave{p_{+}^2}{i} +\ave{p_{-}^2}{i} \cos (2\phi - 2\delta) \right],
\end{split}
\ee
\be
\begin{split}
\delta V_{\rm r,1,\perp}^{(2)} 
=& - \frac{4e \beta_{BA} }{m\hbar} B 
 {\sum_{\beta\neq \alpha}}  \frac{z_{\alpha\beta} \ave{\delta(z)}{\beta\alpha} }{E_{\alpha }-E_{\beta}}  
  \ave{p_{-}^2}{i} \sin (2\phi - 2\delta).
\end{split}
\ee
These are the $c_{11}$ terms in Eqs.~\eqref{eq:gr0} and \eqref{eq:gr0_perp}.

Similarly as before, the quadratic term is zero. For the cubic term, considering $H^\prime_2$ in the first line of Eq.~\eqref{eq:spin_inter} gives
\be
\begin{split}
&\delta H_{\rm r, H_2}^{(2)} 
=  \frac{ e^3 }{m\hbar} (\sigma_x B_x + \sigma_y B_y) (B_x^2+B_y^2) \\
&\times {\sum_{\beta\neq \alpha}}  \frac{  \ave{ \Delta z^2}{\alpha\beta} }{E_{\alpha }-E_{\beta}} \left( \alpha_0 e E_{\rm ext}  z_{\beta\alpha} -\beta_{BA}\ave{ z }{\alpha} \ave{\delta(z)}{\beta\alpha} \right),
\end{split}
\ee
and generates therefore only a parallel component of ${\bf \delta V}$, as the $c_4$, and $c_{12}$ terms in Eq.~\eqref{eq:gr2}. 
Finally, the second and third lines of Eq.~\eqref{eq:spin_inter} result in
\begin{eqnarray}
\delta V_{\rm r,3rd,||}^{(2)}  
&=& - \frac{3 e^3 B^3 }{m^2\hbar} \sum_{\beta\neq\alpha}\sum_{\gamma\neq\alpha}  \frac{z_{\alpha\beta} z_{\beta\gamma} }{(E_{\alpha }-E_{\beta})(E_{\alpha}-E_{\gamma})} \nonumber \\
&& \times \left[ \beta_{BA} \ave{\delta(z)}{\gamma\alpha} \ave{ z }{\alpha} -  \alpha_0 e E_{\rm ext} z_{\gamma \alpha} \right] \nonumber \\
&& \times \left[  \ave{p_{+}^2}{i} - \ave{p_{-}^2}{i} \cos (2\phi - 2\delta) \right] \nonumber \\
&& -  \frac{ \beta_{BA}e^3 B^3 }{m^2\hbar} \sum_{\beta\neq\alpha} \frac{z_{\alpha\beta} \ave{\delta(z)}{\beta\alpha}  \ave{ \Delta z^2 }{\alpha} }{(E_{\alpha }-E_{\beta})^2 } \nonumber \\
&& \times \left[ \ave{p_{+}^2}{i} - \ave{p_{-}^2}{i} \cos (2\phi - 2\delta) \right], \\
\delta V_{\rm r, 3rd,\perp}^{(2)}  
&=& \frac{ \beta_{BA}e^3 B^3 }{m^2\hbar} \sum_{\beta\neq\alpha} \frac{z_{\alpha\beta} \ave{\delta(z)}{\beta\alpha}  \ave{ \Delta z^2 }{\alpha} }{(E_{\alpha }-E_{\beta})^2 } \nonumber \\
&& \times  \ave{p_{-}^2}{i} \sin (2\phi - 2\delta) ,
\end{eqnarray}
entering Eq.~\eqref{eq:gr2} and Eq.~\eqref{eq:gr2_perp} as $c_{7}$, $c_{13}$, and $c_{15}$.

\subsection{Terms from $H_{43}$, $H_{44}$, $H_{45}$, and $H_{47}$ \label{subsec:B-SO}}
The contributions from the in-plane field-induced spin-orbit interaction ($H_{43}$,  $H_{44}$, $H_{45}$ and $H_{47}^{\prime}$) can be computed similarly. Since these terms are directly proportional to the magnetic field, it is more convenient to express them using the Bohr magneton. Namely, we can start with the following expressions,
\begin{eqnarray} 
H_{43} &=& \frac{\lambda_{43}^2}{2\hbar^2} \mu_B ({\bf B} \cdot \boldsymbol{\sigma}) \Big( P_x^2 + P_y^2 + P_z^2 \Big) , \\
H_{44} &=& \frac{\lambda_{44}^2}{4\hbar^2} \mu_B \Big[ \left( \left\{ P_x, P_y \right\} B_y  \sigma_x + \left\{ P_y, P_x \right\} B_x  \sigma_y\right) \nonumber \\
&& \hspace{0.3in} + \left( \left\{ P_z, P_x \right\} B_x + \left\{ P_z, P_y \right\} B_y \right) \sigma_z \Big],\\
H_{45} &=& \frac{\lambda_{45}^2}{2\hbar^2} \mu_B \Big( P_x^2 B_x \sigma_x + P_y^2 B_y \sigma_y \Big).
\end{eqnarray}
For the purpose of this subsection, we also define the following part of $H_{47}$,
\be
H_{47}^{\prime} = -\frac{ \lambda_{47}^{\prime}}{2} \delta (z) \mu_B (B_y \sigma_x + B_x \sigma_y).
\ee
The remaining part of $H_{47}$ is already in the form of a g-tensor,
\be
H_{47}-H_{47}^{\prime} =\frac{e^2\gamma_{47} E_{\rm ext}}{\hbar} (B_y \sigma_x + B_x \sigma_y),
\ee
and therefore does not need a perturbative treatment: it directly gives the term proportional to $\lambda_{47}^3$ in Eq.~\eqref{eq:g470}.

The intra-subband contributions from $H_{43}$ can be put as
\be
\begin{split}
&\delta H_{43}^{(1)} 
= \frac{\lambda_{43}^2}{2\hbar^2} \mu_B ({\bf B} \cdot \boldsymbol{\sigma}) \Big( \ave{p_z^2}{\alpha} + 2 \ave{p_+^2}{i} + e^2 B^2 \ave{\Delta z^2}{\alpha} \Big),
\end{split}
\ee
what gives the $c_{17}$ and $\eta_{+}$ terms in Eq.~\eqref{eq:g430}, and $c_{17}$ and $c_{18}$ terms in Eq.~\eqref{eq:g432}. 
The inter-subband contributions from $H_{43}$ can be written as
\be
\begin{split}
\delta H_{43}^{(2)} 
= &\frac{2 e^2 B^2 \lambda_{43}^2}{m\hbar^2} \mu_B ({\bf B} \cdot \boldsymbol{\sigma}) \Big(  \ave{p_+^2}{i} - \ave{p_-^2}{i} \cos(2\phi-2\delta) \Big) \\
& \times  \sum_{\beta \neq \alpha} \frac{ |z_{\alpha\beta}|^2}{E_{\alpha}-E_{\beta}}  \\
& + \frac{ e^2 B^2 \lambda_{43}^2}{2m\hbar^2} \mu_B ({\bf B} \cdot \boldsymbol{\sigma}) \sum_{\beta \neq \alpha} \frac{ \ave{\Delta z^2}{\alpha\beta} \ave{p_z^2}{\beta\alpha}}{E_{\alpha}-E_{\beta}} \\
& +\frac{3 e^2 B^2 \lambda_{43}^2}{2m^2\hbar^2} \mu_B ({\bf B} \cdot \boldsymbol{\sigma}) \Big(  \ave{p_+^2}{i} - \ave{p_-^2}{i} \cos(2\phi-2\delta) \Big) \\
& \times \sum_{\beta \neq \alpha} \sum_{\gamma \neq \alpha} \frac{z_{\alpha\beta}z_{\beta\gamma} \ave{p_z^2}{\gamma\alpha}}{(E_{\alpha}-E_{\beta}) (E_{\alpha}-E_{\gamma}) }, \\
\end{split}
\ee
what gives the $c_{19}$, $c_{20}$, and $c_{21}$ terms in Eq.~\eqref{eq:g432}. 

The intra-subband contributions from $H_{44}$ can be written as
\be
\begin{split}
\delta H_{44}^{(1)} 
= & \frac{\lambda_{44}^2}{2\hbar^2} \mu_B \Big( B_y \sigma_x + B_x \sigma_y \Big) \ave{p_-^2}{i} \sin(2\delta) \\
& - \frac{\lambda_{44}^2 e^2 B^2 }{4\hbar^2} \mu_B \Big( B_y \sigma_x + B_x \sigma_y \Big) \ave{\Delta z^2}{\alpha} \sin(2\phi) ,
\end{split}
\ee
what gives the $\eta_{-}$ term in Eq.~\eqref{eq:g440} and \eqref{eq:g440_perp}, and $c_{18}$ term in Eq.~\eqref{eq:g442} and \eqref{eq:g442_perp}. 
The inter-subband contributions from $H_{44}$ can be written as
\be
\begin{split}
\delta H_{44}^{(2)} 
= & -\frac{ \lambda_{44}^2 e^2 B^2}{m\hbar^2} \mu_B \Big( B_y \sigma_x + B_x \sigma_y \Big) \sum_{\beta \neq \alpha} \frac{ |z_{\alpha\beta}|^2}{E_{\alpha}-E_{\beta}} \\
 & \times \Big [ \ave{p_+^2}{i} \sin(2\phi) - \ave{p_-^2}{i} \sin(2\delta) \Big],
\end{split}
\ee
what gives the $c_{19}$ term in Eq.~\eqref{eq:g442} and Eq.~\eqref{eq:g442_perp}. 

The intra-subband contributions from $H_{45}$ can be written as
\be
\begin{split}
\delta H_{45}^{(1)} 
= & \frac{ \lambda_{45}^2}{2\hbar^2} \mu_B \Big\{ B_x \sigma_x \Big[ \ave{p_+^2}{i} + \ave{p_-^2}{i} \cos (2 \delta) \Big] \\
& \hspace{0.5in} + B_y \sigma_y \Big[ \ave{p_+^2}{i} - \ave{p_-^2}{i} \cos (2 \delta) \Big] \Big\} \\
& +  \frac{ \lambda_{45}^2 e^2 B^2 }{4\hbar^2} \mu_B \Big( B_y \sigma_x + B_x \sigma_y \Big) \ave{\Delta z^2}{\alpha} \sin(2\phi) ,
\end{split}
\ee
where the first two lines contribute as $\eta_{+}$ and $\eta_{-}$ terms in Eqs.~\eqref{eq:g450} and \eqref{eq:g450_perp}. 
These terms were derived in Ref.~\citenum{alekseev2013}.
The last line enters in Eqs.~\eqref{eq:g452} and \eqref{eq:g452_perp} as $c_{18}$ terms.
The inter-subband contribution from $H_{45}$ is given by
\be
\begin{split}
\delta H_{45}^{(2)} 
= & \frac{\lambda_{45}^2 e^2 B^2}{m\hbar^2} \mu_B \sum_{\beta \neq \alpha} \frac{ |z_{\alpha\beta}|^2}{E_{\alpha}-E_{\beta}} \sin(2\phi)\\
 & \times 
\Big\{ \sigma_x \Big[ \ave{p_+^2}{i} B_y - \ave{p_-^2}{i} B \sin(2\delta-\phi) \Big] \\
& \hspace{0.15in} + \sigma_y \Big[ \ave{p_+^2}{i} B_x - \ave{p_-^2}{i} B \cos(2\delta-\phi) \Big]  \Big\} ,
\end{split}
\ee
what gives the $c_{19}$ term in Eq.~\eqref{eq:g452} and Eq.~\eqref{eq:g452_perp}.

The intra-subband contributions from $H_{47}^{\prime}$ can be written as
\be
\begin{split}
\delta H_{47}^{(1)} 
= &  -\lambda_{47}^{\prime} \ave{\delta (z)}{\alpha} \frac{\mu_B}{2} \Big( B_y \sigma_x + B_x \sigma_y \Big) ,
\end{split}
\ee
giving $c_{22}$ in Eqs.~\eqref{eq:g470} and \eqref{eq:g470_perp}. 
The inter-subband contribution from $H_{47}^{\prime}$ is given by
\be
\begin{split}
\delta H_{47}^{(2)} 
= & -\frac{ \lambda_{47}^{\prime} e^2 B^2 }{2m} \mu_B (B_y \sigma_x + B_x \sigma_y) 
\left\{  \sum_{\beta \neq \alpha} \frac{ \ave{\Delta z^2}{\alpha\beta} \ave{\delta (z)}{\beta\alpha} }{E_{\alpha}-E_{\beta}} \right. \\
& + 3 \sum_{\beta\neq\alpha} \sum_{\gamma\neq\alpha} \frac{ z_{\alpha\beta}  z_{\beta\gamma}  \ave{ \delta(z) }{\gamma\alpha} }{(E_{\alpha }-E_{\beta})(E_{\alpha }-E_{\gamma}) } \\
& \left. \times \left[ \ave{p_{+}^2}{i} - \ave{p_{-}^2}{i} \cos (2\phi - 2\delta) \right] \right\},
\end{split}
\ee
what gives the $c_{23}$ and $c_{24}$ terms in Eqs.~\eqref{eq:g472} and \eqref{eq:g472_perp}.

\subsection{Terms from the inhomogeneous g-factor \label{subsec:interface}}

The important difference to the previously considered spin-orbit interactions, the Zeeman term $H_{\rm Z}$ depends only on the $z$ coordinate, and is therefore diagonal in the in-plane orbital sector of the basis, $\langle i | H_Z |j \rangle \propto \delta_{ij}$. This, first of all, makes the intra-subband contributions zero. For the same reason, in the inter-subband terms, $H^\prime_1$ does not contribute in the first line of Eq.~\eqref{eq:spin_inter}. The only contribution, due to $H^\prime_2$, reads
\begin{eqnarray}
\delta V_{\rm z,H_2,||}^{(2)}  
&=& \frac{\mu_{B} e^2}{2m} B^3 {\sum_{\beta\neq \alpha}}  \frac{ \ave{g(z)}{\alpha\beta} \ave{\Delta z^2}{\beta\alpha} }{E_{\alpha }-E_{\beta}},
\end{eqnarray}
what gives the $c_8$ term in Eq.~\eqref{eq:gi2}.
In the third order of the perturbation theory, the second line of Eq.~\eqref{eq:spin_inter}, we get
\begin{eqnarray}
\delta V_{\rm z,3rd,||}^{(2)}  
&=& \frac{3 e^2 }{m^2} B^3 \sum_{\beta\neq\alpha}\sum_{\gamma\neq\alpha} \frac{z_{\alpha\beta} z_{\beta\gamma}  \ave{g(z)}{\gamma\alpha} }{(E_{\alpha }-E_{\beta})(E_{\alpha}-E_{\gamma})} \nonumber \\
&& \times  \left[  \ave{p_{+}^2}{i} - \ave{p_{-}^2}{i} \cos (2\phi - 2\delta) \right] ,
\end{eqnarray}
what gives the $c_9$ term in Eq.~\eqref{eq:gi2}.

\section{List of all dimensionless constants \label{app:constant}}

\label{app:cs}

In this appendix, we give the correlation factor $\xi_{\alpha i}$ in Eq.~\eqref{eq:xi}, and list all dimensionless constants introduced in Eqs.~\eqref{eq:gd}--\eqref{eq:gi}. The correlation factor is given by
\be
\xi_{\alpha i} = \Phi^2 c_{16} \left[ \eta_+ - \eta_- \cos (2\phi - 2\delta) \right],
\ee
which enters Eqs.~\eqref{eq:gd2}, \eqref{eq:gr2}, \eqref{eq:g432}, \eqref{eq:g472}, \eqref{eq:gd2_perp}, and \eqref{eq:g472_perp} as $c_{16}$.
The dimensionless constants $c_i$ are given by
\begin{subequations}
\begin{flalign} \qquad  c_1&= \frac{\lambda_z}{2 \hbar^2} \ave{\{ \Delta z, p_z^2 \} }{\alpha},&& \end{flalign}
\begin{flalign} \qquad  c_2 &= \frac{2}{m \lambda_z} {\sum_{\beta\neq \alpha}}  \frac{ z_{\alpha\beta} \ave{ p_z^2 }{\beta\alpha} }{E_{\alpha }-E_{\beta}} , &&\end{flalign} 
\begin{flalign} \qquad  c_3 &= \frac{1}{2 \lambda_z^3}  \ave{ \Delta z^3 }{\alpha}, &&\end{flalign} 
\begin{flalign} \qquad  c_4 &= \frac{\hbar^2}{m \lambda_z^5} {\sum_{\beta\neq \alpha}}  \frac{ z_{\alpha\beta} \ave{ \Delta z^2 }{\beta\alpha} }{E_{\alpha }-E_{\beta}} ,&&\end{flalign}
\begin{flalign} \qquad  c_5 &= \frac{1}{2m \lambda_z^3} {\sum_{\beta\neq \alpha}}  \frac{  \ave{\Delta z^2}{\alpha\beta}  \ave{ \{ \Delta z, p_z^2 \}  }{\beta\alpha} }{E_{\alpha }-E_{\beta}} , &&\end{flalign}
\begin{flalign} \qquad  c_6 &=  \frac{3\hbar^2}{m^2 \lambda_z^5}  \sum_{\beta\neq\alpha}\sum_{\gamma\neq\alpha}  \frac{z_{\alpha\beta} z_{\beta\gamma} \ave{ \Delta zp_z^2 }{\gamma\alpha}}{(E_{\alpha }-E_{\beta})(E_{\alpha }-E_{\gamma})},&&\end{flalign} 
\begin{flalign} \qquad  c_7 &=  \frac{3\hbar^4}{m^2 \lambda_z^7} 
 \sum_{\beta\neq\alpha}\sum_{\gamma\neq\alpha}  \frac{z_{\alpha\beta} z_{\beta\gamma} z_{\gamma\alpha} }{(E_{\alpha }-E_{\beta})(E_{\alpha}-E_{\gamma})},&&\end{flalign} 
\begin{flalign} \qquad   c_8 &= \frac{\hbar^2}{m \lambda_z^4} {\sum_{\beta\neq \alpha}}  \frac{ \ave{ g(z) }{\alpha\beta} \ave{ \Delta z^2 }{\beta\alpha}  }{E_{\alpha }-E_{\beta}} , &&\end{flalign} 
\begin{flalign} \qquad  c_9 &=  \frac{3\hbar^4}{m^2 \lambda_z^6} 
 \sum_{\beta\neq\alpha}\sum_{\gamma\neq\alpha}  \frac{z_{\alpha\beta} z_{\beta\gamma} \ave{ g(z) }{\gamma\alpha} }{(E_{\alpha }-E_{\beta})(E_{\alpha}-E_{\gamma})}, &&\end{flalign} 
\begin{flalign} \qquad  c_{10} &= |\psi_{\alpha}(z=0) |^2 \ave{ z }{\alpha},&&\end{flalign}
\begin{flalign} \qquad  c_{11} &= \frac{\hbar^2}{m \lambda_z^2} 
 {\sum_{\beta\neq \alpha}}  \frac{z_{\alpha\beta} \ave{\delta(z)}{\beta\alpha} }{E_{\alpha }-E_{\beta}} , &&\end{flalign} 
\begin{flalign} \qquad  c_{12} &= \frac{\hbar^2}{m \lambda_z^4} 
 {\sum_{\beta\neq \alpha}}  \frac{ \ave{ \Delta z^2 }{\alpha\beta} \ave{ z }{\alpha} \ave{\delta(z)}{\beta\alpha}}{E_{\alpha }-E_{\beta}},&&\end{flalign} 
\begin{flalign} \quad  c_{13} &= \frac{\hbar^4}{m^2 \lambda_z^6} 
 \sum_{\beta\neq\alpha}\sum_{\gamma\neq\alpha}  \frac{ \ave{ z }{\alpha} z_{\alpha\beta} z_{\beta\gamma} \ave{\delta(z)}{\gamma\alpha}}{(E_{\alpha }-E_{\beta})(E_{\alpha}-E_{\gamma})},&&
\end{flalign}
\begin{flalign} \quad  c_{14} &= \frac{\hbar^2}{m^2 \lambda_z^5} \sum_{\beta\neq\alpha}  \frac{z_{\alpha\beta} \ave{ p_z^2 }{\beta\alpha} \ave{ \Delta z^2 }{\alpha}}{(E_{\alpha }-E_{\beta})^2} ,&&
\end{flalign}
\begin{flalign} \quad  c_{15} &= \frac{\hbar^4}{m^2 \lambda_z^6} 
 \sum_{\beta\neq\alpha} \frac{z_{\alpha\beta} \ave{\delta(z)}{\beta\alpha}  \ave{ \Delta z^2 }{\alpha} }{(E_{\alpha }-E_{\beta})^2 },&&
\end{flalign}
\begin{flalign} \quad  c_{16} &= \frac{\hbar^4}{m^2 \lambda_z^6} 
 \sum_{\beta\neq\alpha} \frac{ |z_{\alpha\beta}|^2 }{(E_{\alpha }-E_{\beta})^2 },&&
\end{flalign}
\begin{flalign} \quad  c_{17} &= \frac{ \lambda_z^2 \ave{p_z^2}{\alpha}}{\hbar^2},&&
\end{flalign}
\begin{flalign} \quad  c_{18} &= \frac{ \ave{\Delta z^2}{\alpha} }{ \lambda_z^2 },&&
\end{flalign}
\begin{flalign} \quad  c_{19} &= \frac{\hbar^2}{m \lambda_z^4} 
 \sum_{\beta\neq\alpha} \frac{ |z_{\alpha\beta}|^2   }{E_{\alpha }-E_{\beta} },&&
\end{flalign}
\begin{flalign} \quad  c_{20} &= \frac{1}{m \lambda_z^2} 
 \sum_{\beta\neq\alpha} \frac{ \ave{ \Delta z^2 }{\alpha\beta} \ave{ p_z^2 }{\beta\alpha} }{E_{\alpha }-E_{\beta} },&&
\end{flalign}
\begin{flalign} \quad  c_{21} &= \frac{\hbar^2}{m^2 \lambda_z^4} 
 \sum_{\beta\neq\alpha} \sum_{\gamma\neq\alpha} \frac{ z_{\alpha\beta}  z_{\beta\gamma}  \ave{ p_z^2 }{\gamma\alpha} }{(E_{\alpha }-E_{\beta})(E_{\alpha }-E_{\gamma}) },&&
\end{flalign}
\begin{flalign} \quad c_{22} &= \lambda_z |\psi_{\alpha}(z=0)|^2 , && \end{flalign} 
\begin{flalign} \quad  c_{23} &= \frac{\hbar^2}{m^2 \lambda_z^3} 
 \sum_{\beta\neq\alpha}  \frac{ \ave{ \Delta z^2 }{\alpha\beta}  \ave{ \delta(z) }{\beta\alpha} }{E_{\alpha }-E_{\beta} } , &&
\end{flalign}
\begin{flalign} \quad  c_{24} &= \frac{\hbar^2}{m^2 \lambda_z^5} 
 \sum_{\beta\neq\alpha} \sum_{\gamma\neq\alpha} \frac{ z_{\alpha\beta}  z_{\beta\gamma}  \ave{ \delta(z) }{\gamma\alpha} }{(E_{\alpha }-E_{\beta})(E_{\alpha }-E_{\gamma}) } . &&
\end{flalign}
\end{subequations}

\section{Symmetric quantum well\label{app:symmetric}}

Our main results, Eqs.~\eqref{eq:gd}-\eqref{eq:gz00}, are valid for a general heterostructure potential and therefore also for a symmetric one. However, the latter choice substantially changes the values of constants $c$. Namely, for a symmetric well, only the following constants are nonzero: $c_{11}$, and $c_{15}$-$c_{22}$. With that, the terms $g_{43}$, $g_{44}$, $g_{45}$ and $g_{47,0}$ are the same as given in Eqs.~\eqref{eq:g43}-\eqref{eq:g470}, while the Dresselhaus and the interface term are zero, $g_d=0=g_z$. The remaining terms can be simplified by removing the zero $c$'s. We get
\begin{equation}
g_{\rm r}= \xi_r (4c_{11}-\Phi^2 c_{15} ) [\eta_+ - \eta_- \cos (2\phi-2\delta)],
\label{eq:grHW}
\end{equation}
for the contribution from the Rashba interaction, and 
\begin{equation}
\label{eq:g47HW}
g_{47,2} =  \frac{\lambda_{47}^{\prime}}{\lambda_{z}} c_{22} c_{16} \Phi^2  \sin(2\phi)  
[\eta_+  - \eta_- \cos (2\phi-2\delta)],
\end{equation}
for the $H_{47}$ contribution.

\begin{widetext}

\begin{figure}[h!]
\includegraphics[width=0.49\columnwidth]{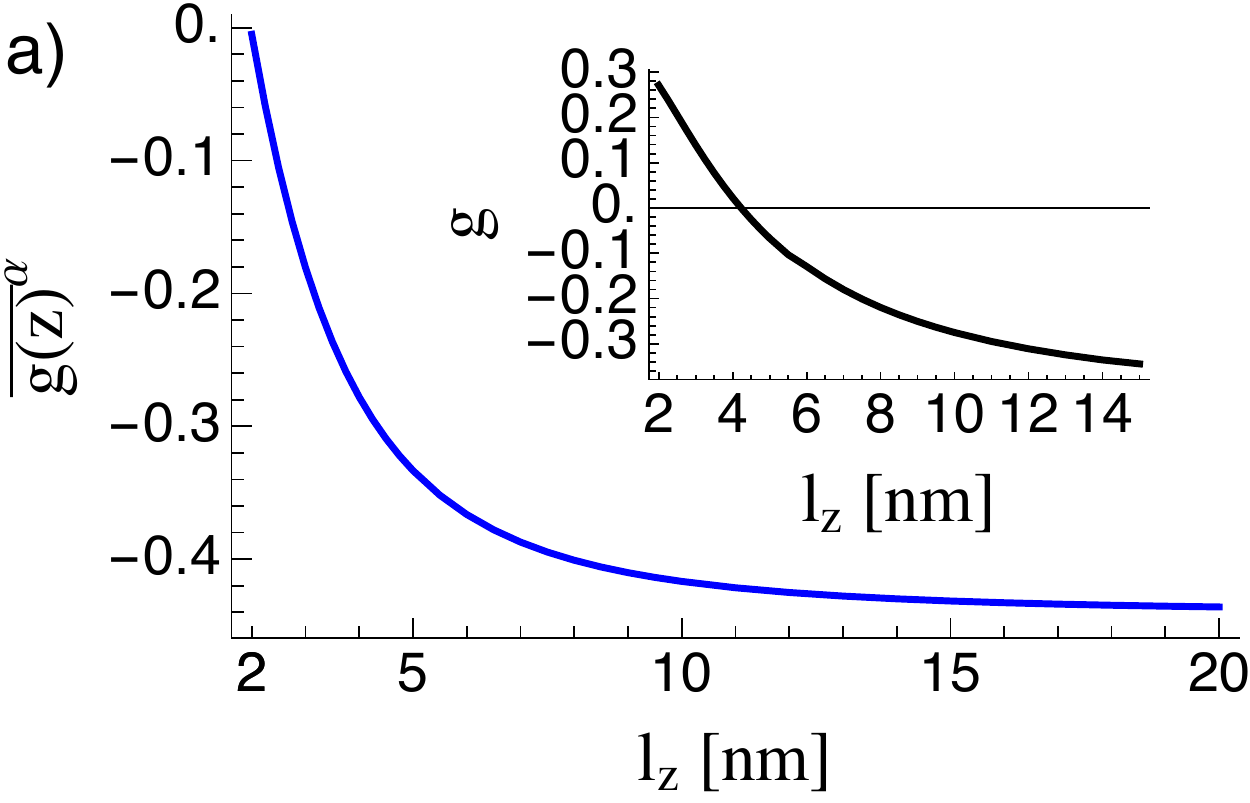}\includegraphics[width=0.49\columnwidth]{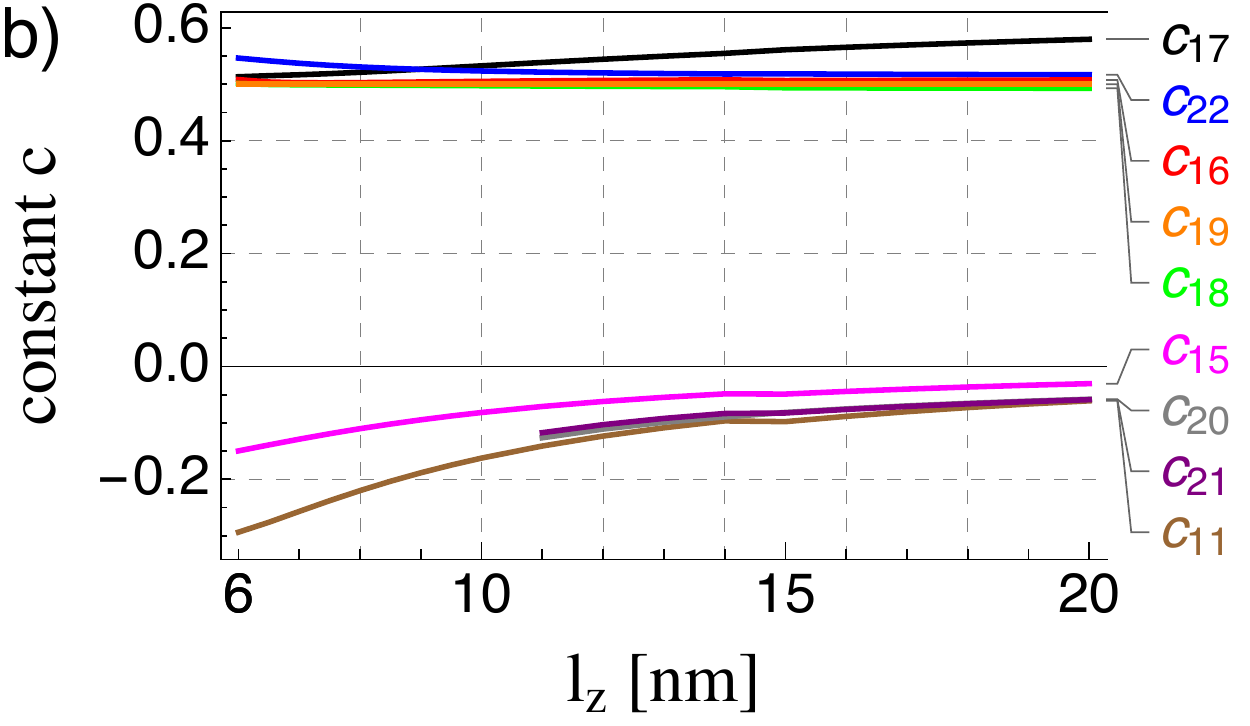}
\\\vspace{0.0cm}
\includegraphics[width=0.49\columnwidth]{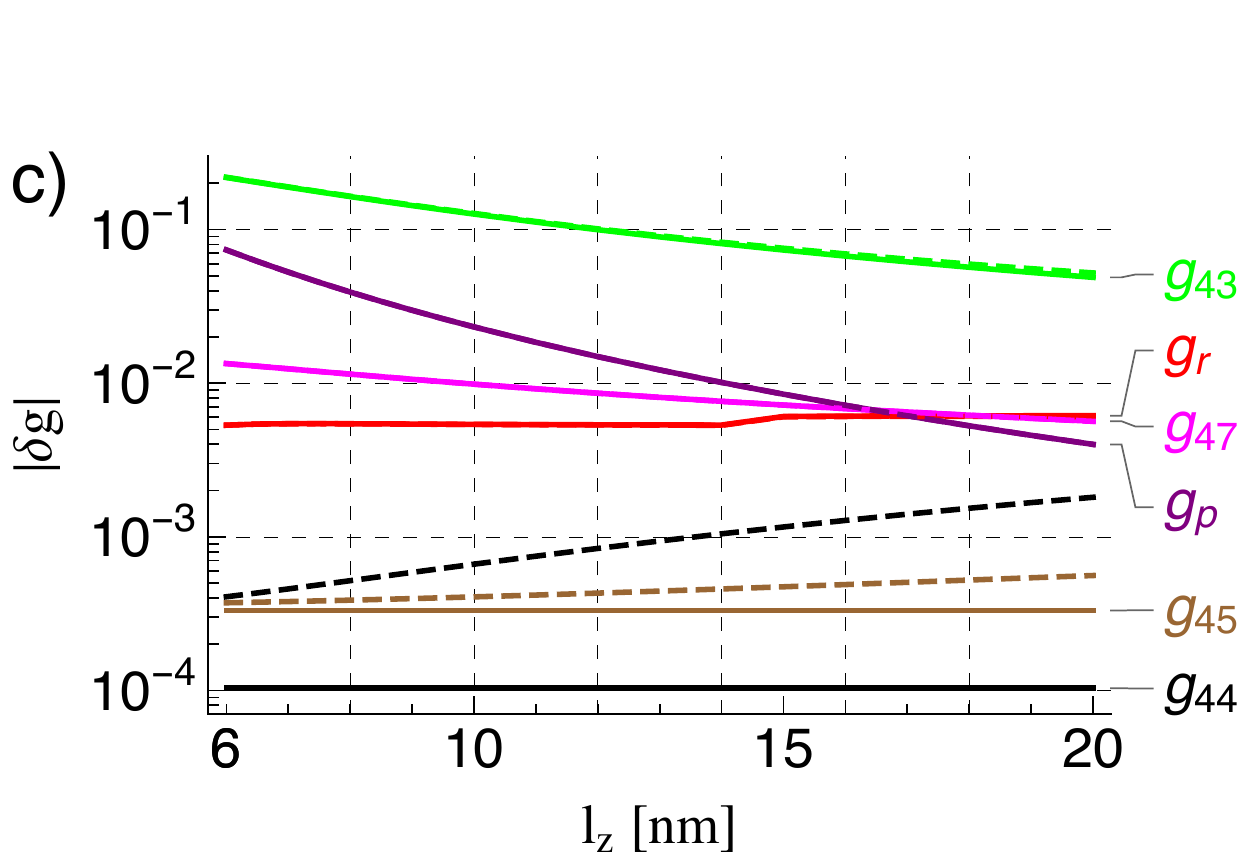}\includegraphics[width=0.49\columnwidth]{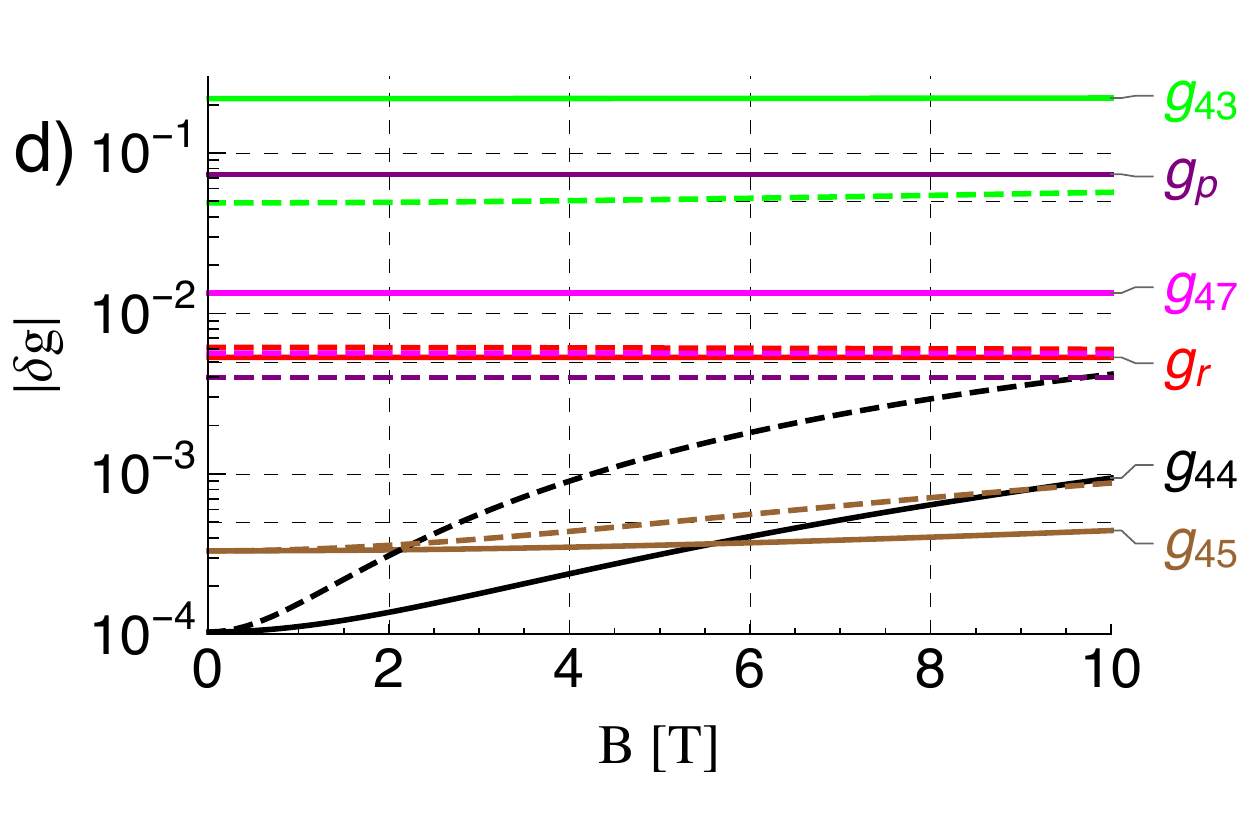}
\\\vspace{0.0cm}
\includegraphics[width=0.49\columnwidth]{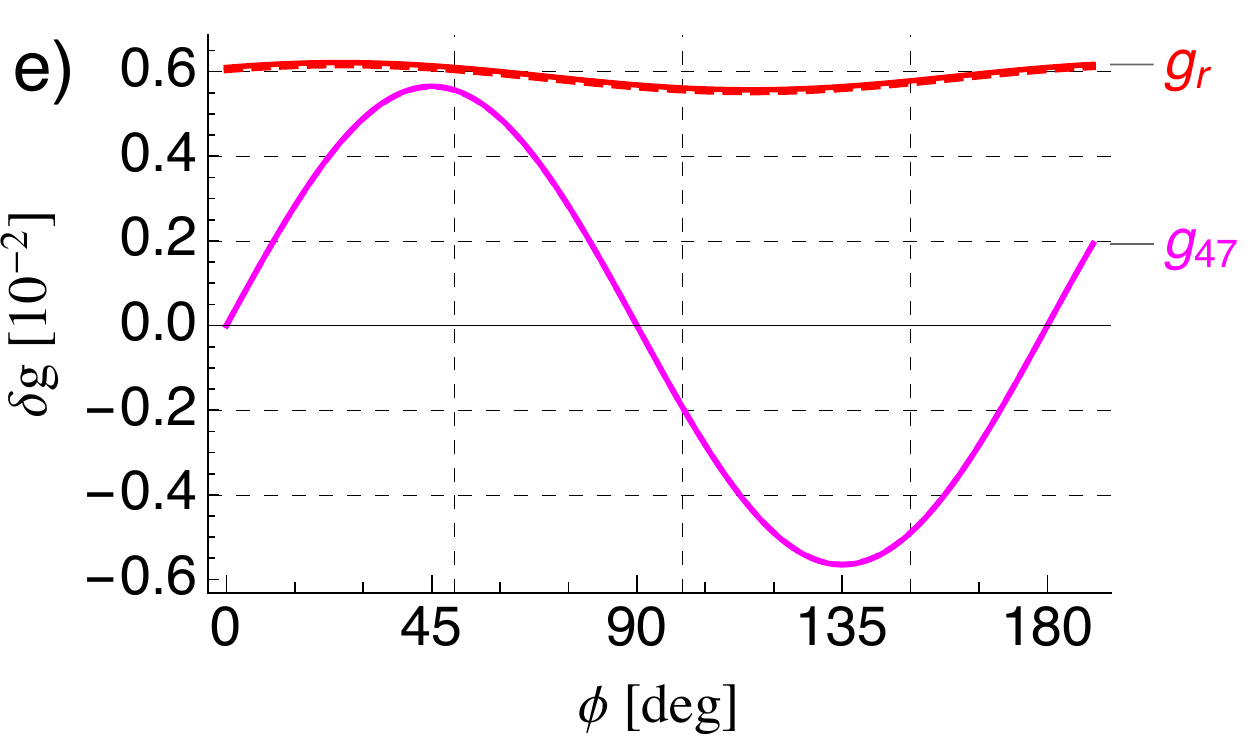}\includegraphics[width=0.49\columnwidth]{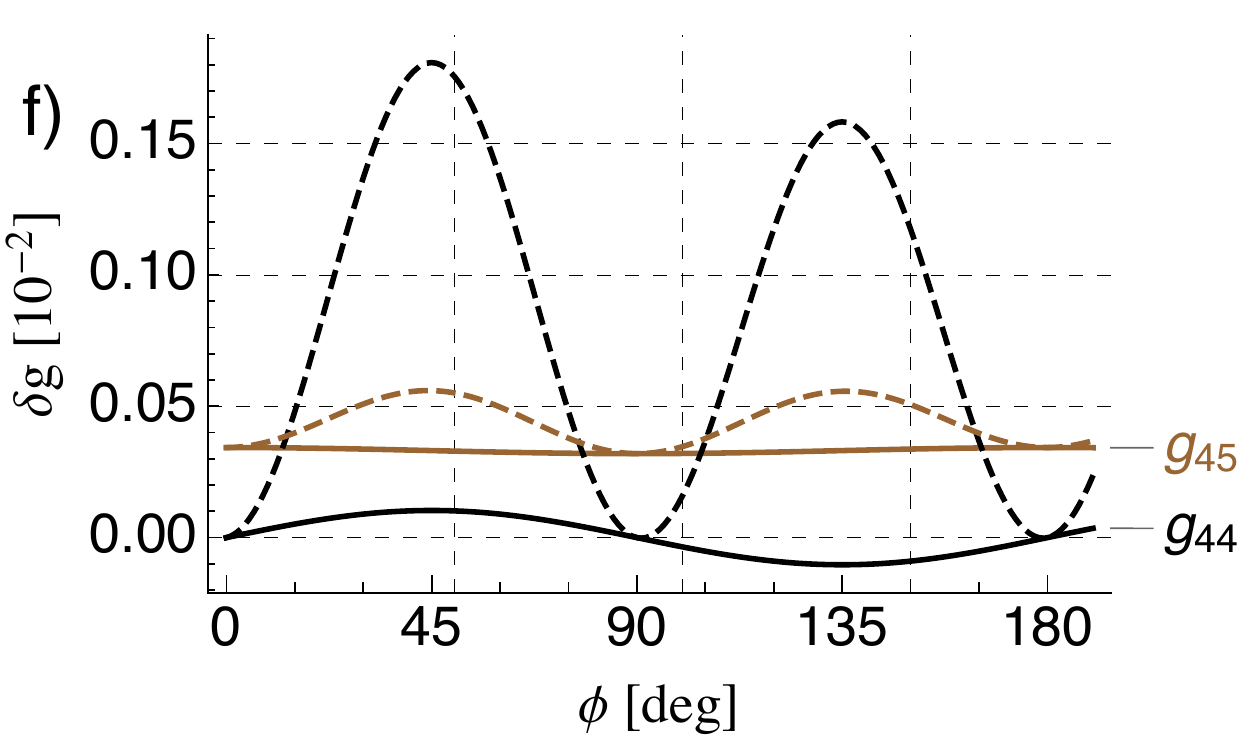}
\\\vspace{0.0cm}
\caption{\label{fig:symmetric}
The corrections to the g-factor, labeled according to the notation of Eqs.~\eqref{eq:gd}--\eqref{eq:gz00}, for a symmetric quantum well with a rectangular confinement. The parameters are the same as in Fig.~\ref{fig:gs} unless stated otherwise. (a) The lowest order approximation to the g-factor, showing the value of $\ave{g(z)}{\alpha}$ for the lowest subband, $\alpha=1$, as a function of the well width, defined in Eq.~\eqref{eq:VR}. Inset: The total g-factor, calculated by adding all corrections to the bulk value in material B.
(b) The values of non-zero constants $c$. Note that $c_{20}$ and $c_{21}$ become formally zero if the quantum well does not have at least 2 subbands. To correct this behavior, one would have to include the contributions from delocalized eigenstates of Eq.~\eqref{eq:Hz}, what we do not do here (the resulting effects on the other panels would be hard to spot).  
(c) Corrections as a function of the quantum well width. The solid (dashed) curves show corrections for $B = 0$ T ($B=6$ T). 
(d) As a function of the magnetic field. The solid (dashed) curves show corrections for $l_z=6$ nm ($l_z=20$ nm). 
(e-f) As a function of the magnetic field orientation. The solid (dashed) curves show corrections for $B=0$ T ($B=6$ T).
}
\end{figure}

\end{widetext}

The above results are valid for a general symmetric quantum-well potential. We now specify to a rectangular potential,
\be
V_z = \left\{
\begin{tabular}{ll}
$V_A$, &if $z \notin \langle -l_z/2, l_z/2 \rangle$,\\
$V_B$, &if $z \in \langle -l_z/2, l_z/2 \rangle$.\\
\end{tabular}
\right.
\label{eq:VR}
\ee
It defines the nominal width as the thickness of the material B layer sandwiched by material A. The effective mass and the g-factor are taken piecewise constant in the three regions. We take the same parameter values for the mass and bulk g-factor in material A and B as given below Eq.~\eqref{eq:Ez}, and use $V_B-V_A=300$ meV. With these amendments, we are ready to analyze the g-factor corrections for a rectangular quantum well quantitatively. 
We plot the g-factor subband average in the main panel of Fig.~\ref{fig:symmetric}(a). Upon narrowing the quantum well, the g-factor grows, reflecting the wavefunction penetrating into AlGaAs. The inset of Fig.~\ref{fig:symmetric}(a) shows the total g-factor as a function of the quantum well width. We plot it to demonstrate the crossing of zero, at about $l_z=4.2$ nm for our parameters, what has been debated some years ago.\cite{ivchenko1992,jeune1997,malinowski2000,arora2013}

We plot the constants $c$ in Fig.~\ref{fig:symmetric}(b). One can see that now they fall into 2 groups, with similar values among their members. Changes in constants $c$ compared to the triangular potential imply changes in the hierarchy of g-factor corrections. Indeed, Fig.~\ref{fig:symmetric}(c)-(d) shows that for a symmetric well, the g-factor correction is basically dominated by a single term, $H_{43}$. For very narrow wells, the penetration might be also visible in experiments with high resolution. On the other hand, there is no appreciable effect from the magnetic field to be expected. Finally, the directional dependence is shown in Fig.~\ref{fig:symmetric}(e)-(f). There is very little variation,\cite{pfeffer2006a} way below the current experimental resolution. The largest variation is from $H_{47}$ and reaches $0.01$, with extrema along the crystal axes.

\section{Magnetic-field-dependent corrections}

\label{app:B2}

\begin{figure}
\includegraphics[width=\columnwidth]{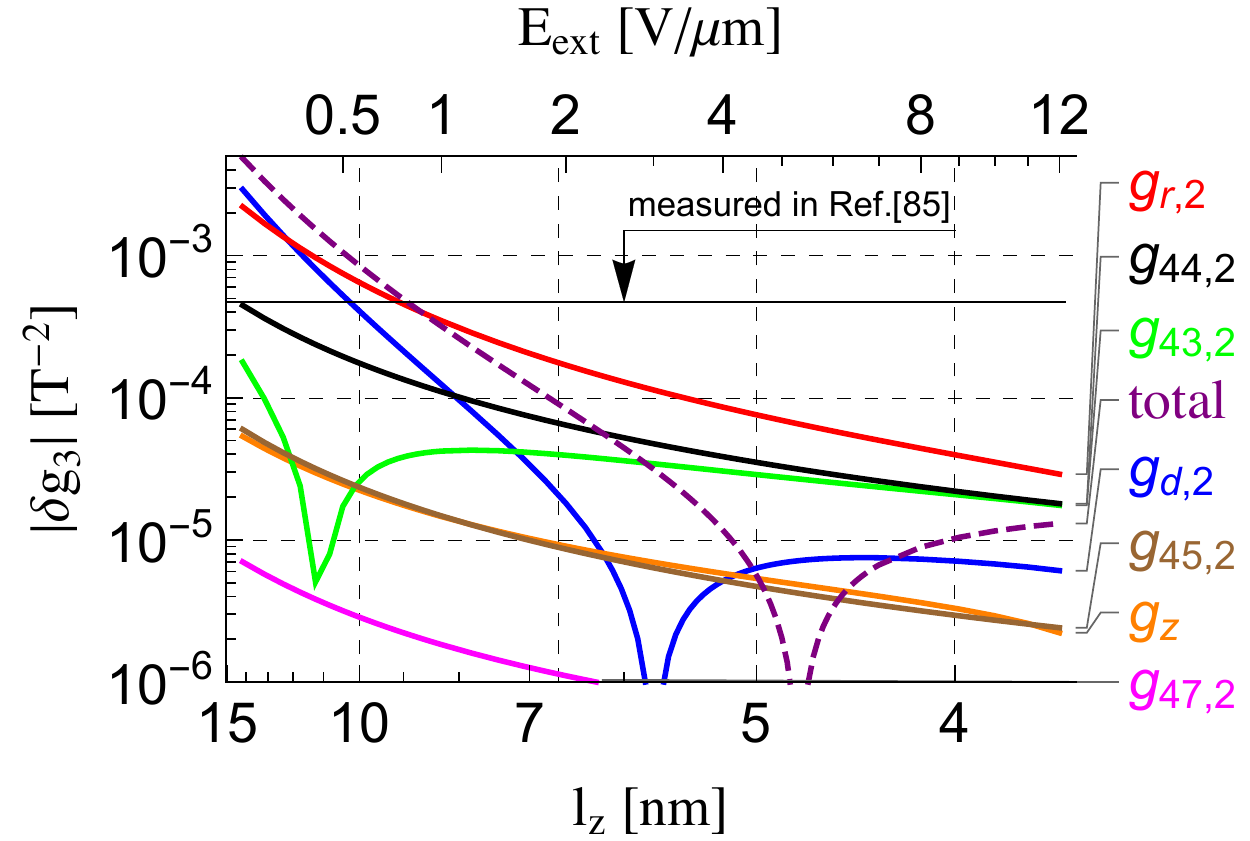}
\caption{\label{fig:cubic}
The g-factor corrections quadratic in the in-plane magnetic field as a function of the 2DEG width. All parameters are the same as in Fig.~\ref{fig:gs}(a) except for $\phi=-45^{\circ}$, chosen in line with the experiment in Ref.~\citenum{fujita2017}. We label the curves using the notation in Eqs.~\eqref{eq:gd}-\eqref{eq:gi} and label the sum of all plotted contributions by ``total''. We transform our dimensionless quantities into the notation of Ref.~\citenum{fujita2017}, defined by $g_3 = g_{\mathrm{x},2} \Binscalar^{-2}$, which has therefore units of T$^{-2}$. In that experiment, the value $g_3 \approx 0.47 \times 10^{-3}$ T$^{-2}$ was fitted from data, which is drawn as a horizontal black solid line. The 2DEG width in that experiment is not known to us.
}
\end{figure}

We obtained corrections to the g-factor which are proportional to the second power of the flux $\Phi$, and therefore second power of the in-plane magnetic field (see also Footnote \ref{fnt:B2}). In the main text, we denoted such terms as $g_{\mathrm{x},2}$, where $\mathrm{x}$ denotes the origin of the term, for example, $\mathrm{x}=\mathrm{d}$ for Dresselhaus. We note that such a cubic Zeeman energy term was fitted from the data measured in Ref.~\citenum{fujita2017}. The notation of that reference, $g_3$, relates to our notation here by $g_3=g_{\mathrm{x},2}\Bin^{-2}$. We plot our results in this notation in Fig.~\ref{fig:cubic}. From that figure, one can see that the Rashba and, for wide 2DEGs, the Dresselhaus terms dominate, respectively. Unlike for the magnetic-field-independent corrections, the term $\mathrm{x}=43$ is not very relevant. We also find an agreement with the value $g_3 \approx +4.7 \times 10^{-4}$ T$^{-2}$ fitted in that experiment, including its sign, for the effective 2DEG width of around $8.5$ nm. Under these conditions, the g-factor B-field nonlinearity is dominated by the Rashba term.

\section{Mixed contribution example}

\label{app:lin-lin}

Here we derive a g-factor correction which is of the second order in spin-orbit couplings. For the sake of illustration, we do it only for the Rashba term, Eq.~\eqref{eq:HR}. According to the scheme followed in Sec.~\ref{sec:2E}, this term would be split according to the powers of the in-plane magnetic field to two terms, $H_{R} = H_{r,0}+H_{r,1}$, with 
\begin{subequations}
\begin{eqnarray}
H_{r,0} &=& \frac{\alpha_R(z)}{\hbar} (p_x \sigma_y - p_y \sigma_x),\\
H_{r,1} &=& \frac{e \alpha_R(z)}{\hbar} (\Ainx  \sigma_y - \Ainy  \sigma_x).
\end{eqnarray}
\end{subequations}
In further, we use the relation
\be
\mathbf{p} = \frac{i m}{\hbar} [h_{2D}, \mathbf{r} ] \equiv \frac{i m}{\hbar} L_{2D}(\mathbf{r}),
\label{eq:iL}
\ee
where the identity sign is the definition of the Liouvillian operator $L_{2D}$ corresponding to the two-dimensional Hamiltonian Eq.~\eqref{eq:H2D}. Since we aim at calculating the contributions to the effective Hamiltonian up to the second order only, we can use the simplified formula given in Eq.~\eqref{eq:Lowdin}. With this, several terms result: choosing the pair $H_{r,n}$-$H_{r,m}$ in the two terms in the latter equation gives four choices, each of which splits to the intra-subband and inter-subband term (eight terms in total).

Let us calculate one of these: the intra-subband contribution coming from a pair $H_{r,0}$-$H_{r,0}$. In this case, using the Liouvillian definition allows us to bring the effective Hamiltonian into the following form
\be
H^{(\alpha)}_{r,0;r,0}(\mathrm{intra}) = \frac{1}{2} [ L^{-1}_{2D}(\ave{H_{r,0}}{\alpha}) , \ave{H_{r,0}}{\alpha} ].
\ee
Using the explicit form of the Liouvillian, the commutator can be evaluated, and we get
\be
H^{(\alpha)}_{r,0;r,0}(\mathrm{intra}) = - m \left(\frac{\ave{\alpha_R(z)}{\alpha}}{\hbar}\right)^2 \left( 1+\sigma_z \frac{L_z}{\hbar}  \right),
\label{eq:grr}
\ee
with $L_z = x p_y - y p_x$. If we now assume a symmetric in-plane confinement potential, $l_x=l_y=l_0$, the expectation value of this operator in the ground state can be easily calculated,
\be
\langle 0 | L_z | 0 \rangle = \frac{e B_z}{4} l_B^2.
\ee
The magnetic-field-renormalized confinement length is
\be
l_B = \left( l_0^{-4} + \frac{e^2 B_z^2}{4\hbar^2} \right)^{-\frac{1}{4}}.
\ee
We can now convert this expression into a renormalization of the out-of-plane component of the g-tensor,
\be
(g_{rr,0})_{zz} = -\frac{1}{2} \left(\frac{\ave{\alpha_R(z)}{\alpha}}{\hbar}\right)^2\frac{m e}{\hbar \mu_B} l_B^2.
\ee
For small out-of-plane fields, so that the magnetic field does not strongly renormalize the confinement, $l_B \approx l_0$, this term evaluates to $(g_{rr,0})_{zz} \approx -0.0012$ for a typical value $l_0=34$ nm. Therefore, it is 1-2 orders of magnitude smaller than the leading terms which are linear in the spin-orbit couplings.

The remaining terms among the possibilities that we enumerated below Eq.~\eqref{eq:iL} are even smaller due to the smallness of various matrix elements of the function $\alpha_R(z)$. For example, the inter-subband term for the pair $H_{r,0}$-$H_{r,0}$ is proportional to the minute value of $\ave{\alpha_R(z)}{\alpha \beta}$. Similarly, the intra-subband term for the pair $H_{r,0}$-$H_{r,1}$ would contain the matrix element $\ave{(z-z_0)\alpha_R(z)}{\alpha}$, etc. All these matrix elements are very small, as they are similar in nature, and in value, to the constants $c_8$ and $c_9$, see App.~\ref{app:cs} and Fig.~\ref{fig:cs}. 

Finally, we note that similar terms would arise from the Dresselhaus interaction. The analogous term, the intra-subband contribution from $H_{d,0}$-$H_{d,0}$, would give
\be
H^{(\alpha)}_{d,0;d,0}(\mathrm{intra}) \approx - m \left(\frac{\gamma_c \ave{p_z^2}{\alpha}}{\hbar^3}\right)^2 \left( 1-\sigma_z \frac{L_z}{\hbar}  \right).
\label{eq:gdd}
\ee
The largest-in-magnitude correction arising in the second order of the spin-orbit coupling is therefore\cite{aleiner2001,stano2005} 
\be
g^{\rm 2nd}_{zz} = \frac{1}{2}\left[\left(\frac{\gamma_c \ave{p_z^2}{\alpha}}{\hbar^3}\right)^2 -  \left(\frac{\ave{\alpha_R(z)}{\alpha}}{\hbar}\right)^2 \right]\frac{m e}{\hbar \mu_B} l_B^2.
\ee
Only the $zz$ component of the g-tensor is changed by a very small value, typically $10^{-3}$. Even though this effect was invoked to interpret the experiment in Ref.~\citenum{koneman2005}, probably a different interaction was responsible for the observed anisotropy there, perhaps $H_{45}$. As we are mostly interested in the in-plane magnetic fields, we do not pursue this issue further.

\end{document}